\documentclass[11pt]{article}
\pdfoutput=1
\usepackage[english]{babel}
\usepackage{amssymb,slashed,latexsym,amsmath,multirow,color}
\usepackage{slashed}
\usepackage{tikz}
\usetikzlibrary{shapes,shadows,arrows}
\usepackage[font=footnotesize,labelsep=newline,labelfont=sc,justification=centering,position=top]{caption}
\usepackage{bookmark}
\usepackage{tikz}
\usetikzlibrary{backgrounds}
\usetikzlibrary{shapes,shadows,arrows}
\tikzstyle{every picture}=[level distance = 8mm, baseline=-0.5ex]
\tikzstyle{prop}=[shape=circle,minimum size=6mm, draw=black!80, fill=green!30]

\pdfoutput=1
\usepackage[font=footnotesize,labelsep=newline,labelfont=sc,justification=centering,position=top]{caption}

\numberwithin{equation}{section}

\usepackage{hyperref}

\newcommand{\ft}[2]{{\textstyle\frac{#1}{#2}}}
\def\rmi{{\rm i}}
\newsavebox{\uuunit}
\sbox{\uuunit}
    {\setlength{\unitlength}{0.825em}
     \begin{picture}(0.6,0.7)
        \thinlines
        \put(0,0){\line(1,0){0.5}}
        \put(0.15,0){\line(0,1){0.7}}
        \put(0.35,0){\line(0,1){0.8}}
       \multiput(0.3,0.8)(-0.04,-0.02){12}{\rule{0.5pt}{0.5pt}}
     \end {picture}}

\newcommand{\U}{\mathop{\rm U}}
\newcommand{\SU}{\mathop{\rm SU}}

\def\be{\begin{equation}}
\def\ee{\end{equation}}
\def\ba{\begin{array}}
\def\ea{\end{array}}
\def\bea{\begin{eqnarray}}
\def\eea{\end{eqnarray}}
\def\bd{\begin{displaymath}}
\def\ed{\end{displaymath}}

\def\nn{\nonumber}

\def\a{\alpha}
\def\b{\beta}
\def\g{\gamma}
\def\G{\Gamma}
\def\d{\delta}

\def\e{\epsilon}

\def\f{\phi}
\def\vf{\varphi}

\def\p{\psi}

\def\l{\lambda}
\def\L{\Lambda}
\def\m{\mu}
\def\n{\nu}
\def\r{\rho}
\def\s{\sigma}
\def\S{\Sigma}
\def\t{\tau}

\def\o{\omega}
\def\O{\Omega}

\def\nn{\nonumber}
\def\cD{\mathcal{D}}
\def\cN{\mathcal{N}}

\def\cL{\mathcal{L}}
\def\cC{\mathcal{C}}

\begin{document}

\begin{flushright}
\hfill{ \
\ \ \ \ MIFPA-13-18\ \ \ \ }
\end{flushright}
\vskip 1.2cm
\begin{center}
{\Large \bf All Off-Shell $R^2$ Invariants in Five Dimensional ${\cal N}=2$ Supergravity
 }
\\

\end{center}
\vspace{25pt}
\begin{center}
{\Large {\bf }}

 \vspace{15pt}

\textbf{Mehmet Ozkan and Yi Pang}

\vspace{20pt}

\centerline{{\small George and Cynthia Woods Mitchell
Institute for Fundamental Physics and Astronomy,}} \centerline{\small Texas
A\&M University, College Station, TX 77843, USA}  \vspace{6pt}

{emails: {\tt mozkan@tamu.edu, pangyi1@physics.tamu.edu}}
\vspace{40pt}

\underline{ABSTRACT}
\end{center}
We construct supersymmetric completions of various curvature squared terms in five dimensional supergravity with eight supercharges. Adopting the dilaton Weyl multiplet, we obtain the minimal off-shell supersymmetric Ricci scalar squared as well as all vector multiplets coupled curvature squared invariants. Since the minimal off-shell supersymmetric Riemann tensor squared and Gauss-Bonnet combination in the dilaton Weyl multiplet have been obtained before, both the minimal off-shell and the vector multiplets coupled curvature squared invariants in the dilaton Weyl multiplet are complete. We also constructed an off-shell Ricci scalar squared invariant utilizing the standard Weyl multiplet. The supersymmetric Ricci scalar squared in the standard Weyl multiplet is coupled to $n$ number of vector multiplets by construction, and it deforms the very special geometry. We found that in the supersymmetric $AdS_5$ vacuum, the very special geometry defined on the moduli space is modified in a simple way. Finally, we study the magnetic string and electric black hole solutions in the presence of supersymmetric Ricci scalar squared.

\vspace{15pt}

\thispagestyle{empty}

\vspace{15pt}

\thispagestyle{empty}

\newpage

\tableofcontents


\newpage


\section{Introduction}
There has been lots of interest in the study of five-dimensional ${\cal N}=2$ supergravity in past years.
On one hand, the solutions in this theory have rich structures including black holes, black rings and
black strings \cite{sab1}-\cite{chong}. On the other hand, this theory can come from string/M theory via Calabi-Yau compactifications \cite{pap1,ant1} which provides a platform for a detailed comparison between the microscopic and macroscopic descriptions of black holes in string theory \cite{Strominger:1996sh,Breckenridge:1996is}. A further compactification of the 5D theory on a circle gives rise to 4D ${\cal N}=2$ supergravity which is important for the study of string triality \cite{Duff:1995sm,Behrndt:1996hu}.

The ordinary classical two-derivative theory does not give a complete description of the physics in the presence of
quantum corrections which can be effectively described by the higher derivative terms. For instance, the five dimensional ${\cal N}=2$ supergravity has gauge-gravity anomaly whose supersymmetrization requires the inclusion of curvature squared terms. Thus, the complete structure of supersymmetric higher derivative terms is important for exploring the full theory at quantum level. Usually, there exist two viewpoints on the supersymmetrization of higher derivative terms. One is the on-shell supersymmetrization, and the other is the off-shell supersymmetrization. Because the off-shell supersymmetrization demands the knowledge of auxiliary fields, currently, it can only be performed up to six dimensions. In string theory, the supersymmetry is on-shell and works only order by order in $\alpha'$. Since the supersymmetry transformation rules depend on $\alpha'$, the supersymmetrization procedure becomes tedious requiring the modification of the transformation rules and the Lagrangian at the same time. However, in the off-shell formalism, supersymmetrizing higher derivative terms can be done without modifying the supersymmetry transformation rules.

In this work, we use five-dimensional superconformal tensor calculus \cite{Bergshoeff:2001hc}-\cite{Coomans:2012cf} which is an off-shell formalism facilitating the construction tremendously\footnote{A superspace formulation of five dimensional N=2 supergravity and matter multiplets has been obtained in \cite{Kuzenko:2007hu}.}. In five dimensions, there are two inequivalent Weyl multiplets: the standard Weyl multiplet and the dilaton Weyl multiplet. The main difference between these two Weyl multiplets is that the dilaton Weyl multiplet contains a graviphoton in its field content whereas the standard Weyl multiplet does not. A supergravity theory based on the standard Weyl multiplet requires coupling to an external vector multiplet.

Utilizing the standard Weyl multiplet, the supersymmetric Weyl tensor squared invariant has been found in
\cite{Hanaki:2006pj} while the supersymmetric Riemann squared \cite{Bergshoeff:2011xn, Ozkan:2013uk} and Gauss-Bonnet combination \cite{Ozkan:2013uk} are based on the dilaton Weyl multiplet. Similar constructions in $D=4$, ${\cal N}=2$ and $D=6,\,\cN=(1,0)$ theories have been done in \cite{LopesCardoso:2000qm, deWit:2006gn, butt,Bergshoeff:1986vy}. Therefore, for the completeness of off-shell curvature squared invariants, the missing supersymmetric curvature squared terms in $D=5,\, {\cal N}=2$ theory are the Riemann tensor squared in the standard Weyl multiplet and Ricci scalar squared in both Weyl multiplets. Using the vector multiplet actions, we derive the supersymmetric completion of Ricci scalar squared term by composing the fields in the vector multiplet in terms of the elements of linear multiplet and Weyl multiplet. As a consequence, the supersymmetric curvature squared structures become complete in the dilaton Weyl multiplet. We then proceed to couple all these curvature squared invariants to $n$ number of vector multiplets. When the standard Weyl multiplet is adopted, the Ricci scalar squared term is coupled to $n$ Abelian vector multiplets. We show that it modifies the very special geometry defined on the moduli space of the $n$ vector multiplets. Compared with the supersymmetric Weyl tensor squared action \cite{Hanaki:2006pj}, the supersymmetric Ricci scalar squared term provides a simpler modification to the two-derivative theory. Although, the Ricci scalar squared can be generated by a field redefinition from the string theory viewpoint, the off-shell supersymmetric completion of Ricci scalar squared term is an independent worthwhile superinvariant and does modify the physics in the context of gauge/gravity duality \cite{Blau:1999vz,Nojiri:1999mh}.

The remainder of this paper is organized as follows. In section \ref{section: multiplets}, we introduce the superconformal multiplets in $D=5$, ${\cal N}=2$ superconformal theory. In section \ref{section: sactions}, we list the superconformal actions for the matter multiplets. In section \ref{section: alldilaton} we first review the previously constructed minimal off-shell curvature squared invariants purely based on the dilaton Weyl multiplet including the supersymmetric Riemann squared action and the Gauss-Bonnet action. Then, we construct the minimal off-shell Ricci scalar squared invariant. In section \ref{section: dilationvector}, we derive the vector multiplets coupled Rieman tensor squared and Ricci scalar squared invariants and review the vector multiplets coupled Weyl tensor squared for a complete discussion. Starting from section \ref{section: Sugra} we begin to use the standard Weyl multiplet. We obtain an off-shell two-derivative Poincar\'e supergravity by using the linear and vector multiplets as compensators and  gauge the $\U(1)$ $R$-symmetry by coupling this theory to $n$ number of vector multiplets. In section \ref{section: Ricci2SW}, after a brief discussion about the supersymmetric Weyl tensor squared, we construct an off-shell vector multiplets coupled Ricci scalar squared invariant. We then analyze the effects of the Ricci scalar squared term to very special geometry, particularly in the case of $AdS_5$ vacuum. In section \ref{section: Solution}, we study the supersymmeric magnetic string and electric black hole solutions. We summarize in section \ref{section: conc}.


\section{Multiplets of Five Dimensional Superconformal Theory}\label{section: multiplets}

In this section, we introduce the five dimensional $\cN=2$ superconformal multiplets to be used in the construction of off-shell two-derivative Poincar\'e supergravities and curvature squared actions. The section starts with the introduction of two superconformal Weyl multiplets. A superconformal Weyl multiplet contains all the gauge fields associated with the superconformal algebra as well as proper matter fields. The latter is included in order to balance the bosonic and fermionic degrees of freedom and implement the off-shell closure of the algebra. In $D=5,\, \cN=2$ theory, there exist two different choices for the matter fields which leads to two different Weyl multiplets: the standard Weyl multiplet and the dilaton Weyl multiplet. In the first two subsections, we exhibit the standard Weyl multiplet and the dilaton Weyl multiplet. The last two subsections are devoted to the review of the superconformal matter multiplets including the vector multiplet and the linear multiplet which will be used as the compensating multiplets in the construction of superconformal actions.


\subsection{The Standard Weyl Multiplet}\label{ss: standardweyl}

The standard Weyl multiplet contains 32+32 off-shell degrees of freedom including a f\"unfbein $e_\m{}^a$, a gravitino, $\p_\m^i$, a dilaton gauge field $b_\m$, an $\SU(2)$ gauge field $V_\m{}^{ij}$, a scalar $D$, an antisymmetric tensor $T_{ab}$, and a symplectic Majorana spinor $\chi^i$. The full $Q$, $S$ and $K$ transformations of these fields are given by\footnote{In this paper, we use the conventions of \cite{Bergshoeff:2001hc} where the signature of the metric is diag$(-, +, +, +, +)$. A table introducing the correspondence between the notations of \cite{Hanaki:2006pj} and \cite{Bergshoeff:2001hc} is given in Appendix \ref{section: notation} for reader's convenience.} \cite{Bergshoeff:2001hc}
\bea
\d e_\m{}^a   &=&  \ft 12\bar\e \g^a \psi_\m  \nn\, ,\\
\d \psi_\m^i   &=& (\partial_\m+\tfrac{1}{2}b_\m+\tfrac{1}{4}\omega_\m{}^{ab}\g_{ab})\e^i-V_\m^{ij}\e_j + \rmi \g\cdot T \g_\m
\e^i - \rmi \g_\m
\eta^i  \nn\, ,\\
\d V_\m{}^{ij} &=&  -\ft32\rmi \bar\e^{(i} \phi_\m^{j)} +4
\bar\e^{(i}\g_\m \chi^{j)}
  + \rmi \bar\e^{(i} \g\cdot T \psi_\m^{j)} + \ft32\rmi
\bar\eta^{(i}\psi_\m^{j)} \nn\, ,\\
\d T_{ab} &=& \tfrac12 \rmi\bar\e \g_{ab} \chi - \tfrac3{32} \rmi \bar\e \widehat{R}_{ab}(Q)\,, \nn
\eea
\bea
\d \chi^i &=& \tfrac14 \e^i D - \tfrac1{64} \g \cdot  \widehat{R}^{ij}(V) \e_j + \tfrac18 \rmi \g^{ab}\slashed{\mathcal{D}}T_{ab}\e^i - \tfrac18 \rmi \g^a  \mathcal{D}^b T_{ab} \e^i \nn\\
&& - \tfrac14 \g^{abcd}T_{ab} T_{cd} \e^i + \tfrac16 T^2 \e^i + \tfrac14 \g \cdot T \eta^i\,, \nn\\
\d D &=& \bar\e \slashed{\mathcal{D}}\chi - \tfrac53 \rmi \bar\e \g \cdot T \chi - \rmi \bar\eta \chi\,, \nn\\
 \d b_\m       &=& \ft12 \rmi \bar\e\phi_\m -2 \bar\e\g_\mu \chi +
\ft12\rmi \bar\eta\psi_\mu+2\Lambda _{K\mu } \,,
\label{SWMTR}
\eea
where
\bea
\mathcal{D}_\m\chi^i&=&(\partial_\mu - \tfrac72 b_\mu +\tfrac14 \omega_\mu{}^{ab} \g_{ab})\chi^i -V_\mu^{ij}\chi_j
 - \tfrac14 \p_\m^i D  + \tfrac1{64} \g \cdot  \widehat{R}^{ij}(V) \p_{\m j}\nn\\
&& - \tfrac18 \rmi \g^{ab}\slashed{\mathcal{D}}T_{ab}\p_\m^i + \tfrac18 \rmi \g^a  \mathcal{D}^b T_{ab} \p_\m^i  + \tfrac14 \g^{abcd}T_{ab} T_{cd} \p_\m^i - \tfrac16 T^2 \p_\m^i - \tfrac14 \g \cdot T \phi_\m^i\,, \nn \\
\mathcal{D}_\m T_{ab}&=&\partial_\m T_{ab}-b_\m T_{ab}-2\omega_\mu{}^c{}_{[a}T_{b]c}-\tfrac{1}{2}i\bar{\p}_\m\g_{ab}\chi+\tfrac{3}{32}i\bar{\p}_\m\widehat{R}_{ab}(Q)\,.
\eea
The supercovariant curvatures appearing in the transformation rules (\ref{SWMTR}) are given by \footnote{A more detailed discussion about the supercovariant curvatures can be found in \cite{Bergshoeff:2001hc}}
\bea
\widehat{R}_{\m\n}{}^{ab}(M)&=&2\partial_{[\m}\o_{\n ]}{}^{ab}+2\o_{[\m}{}^{ac}\o_{\n ]c}{}^{b} + 8 f_{[\m}{}^{[a}e_{\n ]}{}^{b]}+\rmi \bar\p_{[\m}\g^{ab}\p_{\n ]} + \rmi \bar\p_{[\m}\g^{[a} \g \cdot T \g^{b]}\p_{\n ]}  \nn\\
&& +\bar\p_{[\m} \g^{[a} \widehat{R}_{\n ]}{}^{b]}(Q)+\tfrac12 \bar\p_{[\m}\g_{\n ]} \widehat{R}^{ab}(Q) -8 \bar\p_{[\m} e_{\n ]}{}^{[a} \g^{b]}\chi+i\bar{\phi}_{[\m} \g^{ab} \p_{\n]}\nn\,, \\
\widehat{R}_{\m\n}{}^{ij}(V)&=&2\partial_{[\m} V_{\n]}{}^{ij} -2V_{[\m}{}^{k( i}
V_{\n ]\,k}{}^{j)} {-3\rmi}{\bar\f}^{( i}_{[\m}\p^{j)}_{\n ]}  - 8 \bar{\p}^{(i}_{[\m}
\g_{\n]} \chi^{j)} - \rmi \bar{\p}^{(i}_{[\m} \g\cdot T \psi_{\n]}^{j)}   \,, \\
\widehat{R}_{\m\n}^i(Q)&=&2\partial_{[\m}\p_{\n]}^i+\frac{1}{2}\o_{[\m}{}^{ab}\g_{ab}\p_{\n]}^i+b_{[\m}\p_{\n]}^i-2V_{[\m}^{ij}\p_{\n] j}-2i\g_{[\m}\phi_{\n]}^i+2i\g\cdot T\g_{[\m}\p_{\n]}^i\,, \nn
\eea
where the spin connection $\omega_{\m}{}^{ab}$, the $S$-supersymmetry gauge field $\phi_\m^i$ and the special conformal symmetry gauge field $f_\m{}^a$ are composites. Their explicit expressions are given by
 \bea \o_\m{}^{ab}
&=& 2 e^{\n[a} \partial_{[\m} e_{\n]}^{~b]} - e^{\n[a} e^{b]\s} e_{\m c}
\partial_\n e^{~c}_\s
 + 2 e_\m^{~~[a} b^{b]} - \ft12 \bar{\p}^{[b} \g^{a]} \p_\m - \ft14
\bar{\p}^b \g_\m \p^a \,,\nonumber\\
\f^i_\m &=& \ft13\rmi \g^a \widehat{R}^\prime _{\m a}{}^i(Q) - \ft1{24}\rmi
\g_\m \g^{ab} \widehat{R}^\prime _{ab}{}^i(Q)\,, \label{transfDepF} \\
f^a_\m &=&  - \ft16{\cal R}_\mu {}^a + \ft1{48}e_\mu {}^a {\cal R},\quad {\cal R}_{\mu \nu }\equiv \widehat{R}_{\mu \rho }^{\prime~~ab}(M) e_b{}^\rho
e_{\nu a},\quad {\cal R}\equiv {\cal R}_\mu {}^\mu\,, \nonumber
\label{cf1}
\eea
where we used the notation $\widehat{R}_{ab}'(Q)$ and $\widehat{R}_{\mu \rho }^{\prime~~ab}(M)$ to indicate that these expressions are obtained from $\widehat{R}_{ab}(Q)$ and $\widehat{R}_{\mu \rho }^{~~ab}(M) $ by omitting the $\phi_\m^i$ and $f_{\m}^a$ terms respectively.


\subsection{The Dilaton Weyl Multiplet}\label{ss: dilatonWeyl}

The gauge sector of the dilaton Weyl multiplet is the same as the gauge sector of the standard Weyl multiplet. The matter sector of the dilaton Weyl multiplet differs from that of the standard Weyl as it contains a physical vector $C_\m$, an antisymmetric two-form gauge field $B_{\m\n}$, a dilaton field $\s$ and a dilatino $\p^i$. The $Q$-, $S$- and $K$- transformation rules of the fields in the dilaton Weyl multiplet can be found in \cite{Bergshoeff:2001hc}
\bea
\d e_\m{}^a   &=&  \ft 12\bar\e \g^a \psi_\m  \nn\, ,\\
\d \psi_\m^i   &=& (\partial_\m+\tfrac{1}{2}b_\m+\tfrac{1}{4}\omega_\m{}^{ab}\g_{ab})\e^i-V_\m^{ij}\e_j + \rmi \g\cdot \underline{T} \g_\m
\e^i - \rmi \g_\m
\eta^i  \nn\, ,\\
\d V_\m{}^{ij} &=&  -\ft32\rmi \bar\e^{(i} \phi_\m^{j)} +4
\bar\e^{(i}\g_\m \underline{\chi}^{j)}
  + \rmi \bar\e^{(i} \g\cdot \underline{T} \psi_\m^{j)} + \ft32\rmi
\bar\eta^{(i}\psi_\m^{j)} \,, \nn\\
 \d C_\m
&=& -\ft12\rmi \s \bar{\e} \p_\m + \ft12
\bar{\e} \g_\m \p, \nn\\
 \d B_{\m\n}
&=& \ft12 \s^2 \bar{\e} \g_{[\m} \p_{\n]} + \ft12 \rmi \s \bar{\e}
\g_{\m\n} \p + C_{[\m} \d(\e) C_{\n]}, \nonumber
\eea
\bea
\d \p^i &=& - \ft14 \g \cdot \widehat{G} \e^i -\ft12\rmi \slashed{\mathcal{D}} \s
\e^i + \s \g \cdot \underline{T} \e^i -\ft14\rmi\s^{-1}\e_j \bar\p^i \p^j  + \s\eta^i \,,\nonumber\\
\d \s &=& \ft12 \rmi \bar{\e} \p \, ,\nonumber\\
 \d b_\m       &=& \ft12 \rmi \bar\e\phi_\m -2 \bar\e\g_\mu \underline{\chi} +
\ft12\rmi \bar\eta\psi_\mu+2\Lambda _{K\mu } \,,
\label{TransDW}
\eea
where
\begin{eqnarray}
\mathcal{D}_\mu\, \s &=&  (\partial_\mu - b_\mu) \s
- \tfrac12\, \rmi\bar{\psi}_\mu \p \ ,
\nn\\
\mathcal{D}_\mu \p^i &=&  (\partial_\mu -\ft32 b_\mu +\ft14\,  \o_\mu{}^{ab}\g_{ab} ) \p^{i} - V_\mu^{ij} \p_j +\tfrac 14 \g
\cdot \widehat{G} \p_\mu^i  \nn\\
&& + \ft12\rmi \slashed{\cD} \s \p_\mu^i
 +\ft14\rmi\s^{-1}\p_{\m j}\bar\p^i\p^j  - \s  \g \cdot \underline{T} \p_\mu^i - \s
\phi_\mu^i\,,
\label{cd1}
\end{eqnarray}
and the supercovariant curvatures are defined according to
\bea
\widehat{G}_{\m\n}  &=& G_{\m\n} - \bar{\p}_{[\m} \g_{\n]} \p + \tfrac 12 \rmi
\s \bar{\p}_{[\m} \p_{\n]} \label{hatG} ,\nn\\
\widehat{H}_{\m\n\r} &=& H_{\m\n\r} - \ft34
\s^2 \bar{\p}_{[\m} \g_\n \p_{\r]} - \ft32\rmi \s \bar{\p}_{[\m}
\g_{\n\r]} \p.
\label{DefH}
\eea
In the above expressions, $G_{\m\n}=2 \partial_{[\mu } C_{\nu ]}$ and $H_{\m\n\r} = 3\partial _{[\mu }B_{\nu \rho ]} + \ft32 C_{[\m}  G_{\n\r]}$. Note that $\widehat{G}_{\m\n}$ and ${\widehat H}_{\mu\nu\rho}$ are invariant under the following gauge transformations
\be \delta C_\mu= \partial_\mu\Lambda\ ,\qquad \delta B_{\mu\nu} =
2\partial_{[\mu} \Lambda_{\nu]} -\ft12 \Lambda G_{\mu\nu}.
\ee
The underlined expressions $\underline{T}^{ab}, \underline{\chi}^i$ and $\underline{D}$, which are the fundamental auxiliary fields in the standard Weyl multiplet, become composite expressions in the dilaton Weyl multiplet \cite{Bergshoeff:2001hc}
\bea
\underline{T}^{ab} &=& \ft18 \s^{-2} \Big( \s \widehat G^{ab} + \ft16 \e^{abcde} \widehat H_{cde} + \ft14 \rmi \bar\p \g^{ab} \p \Big) \,, \nn\\
\underline{\chi}^i &=& \ft18 \rmi \s^{-1} \slashed{\cD} \p^i + \ft1{16} \rmi \s^{-2} \slashed{\cD} \s \p^i - \ft1{32} \s^{-2} \g \cdot \widehat G \p^i + \ft14 \s^{-1} \g \cdot \underline{T} \p^i \nn\\
&& + \ft1{32} \rmi \s^{-3} \p_j \bar\p^i\p^j ,\,\nn\\
\underline{D} &=& \ft14 \s^{-1} \Box^c \s + \ft18 \s^{-2} (\cD_a \s) (\cD^a \s) - \ft1{16} \s^{-2} \widehat G_{\m\n} \widehat G^{\m\n}\nn\\
&& - \ft18 \s^{-2} \bar\p \slashed{\cD} \p  -\ft1{64} \s^{-4} \bar\p^i \p^j \bar\p_i \p_j - 4 \rmi \s^{-1} \p \underline{\chi} \nn\\
&& + \Big( - \ft{26}3 \underline{T_{ab}} + 2 \s^{-1} \widehat G_{ab} + \ft14 \rmi \s^{-2} \bar\p \g_{ab} \p \Big) \underline{T}^{ab}\,,
\label{UMap}
\eea
where the superconformal d'Alambertian for $\s$ is given by
\bea
&&\Box^c \s= (\partial^a - 2b^a + \o_b{}^{ba}) \cD_a \s - \ft12 \rmi \bar\p_a \cD^a\psi - 2\sigma \bar\p_a \g^a {\chi} \nn\\
&&\quad\quad + \ft12 \bar\p_a \g^a \g \cdot \underline{T} \psi + \ft12 \bar\phi_a \g^a \psi + 2 f_a{}^a \sigma \,.
\eea
These composite expressions define a map from the dilaton Weyl multiplet to the standard Weyl multiplet.

\subsection{The Vector Multiplet} \label{ss: vector}
The off-shell Abelian $D=5$, $\mathcal{N}=2$ vector multiplet
contains $8+8$ degrees of freedom. Its bosonic sector consists of a vector field $A_\m$, a scalar field $\r$ and
an auxiliary $\SU(2)$ triplet field $Y^{ij} = Y^{(ij)}$. The fermionic sector contains an $\SU(2)$ doublet
$\l^i$. The $Q$- and $S$-transformations for the vector multiplet are given by \cite{Bergshoeff:2002qk}
\begin{eqnarray}
\d A_\m &=& -\ft12\rmi \r \bar{\e} \p_\m + \ft12 \bar{\e}
\g_\m \l \ ,
\nn\\
\d Y^{ij} &=& -\ft12
\bar{\e}^{(i} \slashed{\mathcal{D}} \l^{j)} + \ft12 \rmi \bar{\e}^{(i}
\g \cdot T \l^{j)} - 4 \rmi \s \bar{\e}^{(i} \chi^{j)} +
\ft12 \rmi \bar{\eta}^{(i} \l^{j)}\,, \nn\\
\d\l^{i} &=& - \ft14 \g \cdot \widehat{F} \e^i -\ft12\rmi
\slashed{\mathcal{D}}\r \e^i + \r \g \cdot T \e^i - Y^{ij} \e_j +
\r \eta^i \ ,
\nn\\
\d\r &=& \ft12 \rmi \bar{\e}
\l . \label{VMTR}
\end{eqnarray}
In the above expressions, the superconformally covariant derivatives are defined as
\begin{eqnarray}
\mathcal{D}_\mu\, \r &=&  (\partial_\mu - b_\mu) \r
- \tfrac12\, \rmi\bar{\psi}_\mu \l \ ,
\nn\\
\mathcal{D}_\mu \l^i &=&  (\partial_\mu -\ft32 b_\mu +\ft14\,  \o_\mu{}^{ab}\g_{ab} ) \l^{i} - V_\mu^{ij} \l_j \nn\\
&& +\tfrac 14 \g
\cdot \widehat{F} \p_\mu^i + \ft12\rmi \slashed{\mathcal{D}} \r \p_\mu^i
 + Y^{ij} \p_{\mu\, j} - \r  \g \cdot T \p_\mu^i - \r
\phi_\mu^i,
\label{cd2}
\end{eqnarray}
where the supercovariant Yang-Mills curvature is defined as
\begin{equation}
\widehat{F}_{\m\n}  = F_{\m\n} - \bar{\p}_{[\m} \g_{\n]} \l + \tfrac 12 \rmi
\r \bar{\p}_{[\m} \p_{\n]}\,,\quad F_{\m\n}=2 \partial_{[\mu } A_{\nu ]}.
\label{hatF}
\end{equation}
The local supersymmetry transformation rules given in (\ref{VMTR}) are obtained by coupling the rigid supersymmetric theory to a Weyl multiplet \cite{Bergshoeff:2002qk}. In the above transformation rules, we utilized the standard Weyl multiplet. If the dilaton Weyl multiplet is considered, the supersymmetry tranformation rules can be obtained straightforwardly by replacing $T_{ab}, D$ and $\chi^i$ by their composite expressions according to (\ref{UMap}).

\subsection{The Linear Multiplet} \label{ss: linear}
The off-shell $D=5, \, {\cal{N}} = 2$ linear multiplet contains $8+8$ degrees of freedom. The bosonic fields consist of an $\SU(2)$ triplet $L^{ij} = L^{(ij)}$, a constrained vector $E_{a}$ and a scalar $N$. The fermionic field contains an $\SU(2)$ doublet
$\varphi^i$. Adopting the standard Weyl multiplet, the $Q$-and $S$-supersymmetry transformations of the linear multiplet are given in \cite{Bergshoeff:2002qk}
\begin{eqnarray}
\delta L^{ij} &=& \rmi \bar{\e}^{(i}\varphi^{j)}\,,\nn\\
\delta \varphi^{i} &=& - \tfrac{1}{2} \rmi \slashed{\mathcal{D}}L^{ij}\e_{j} - \tfrac{1}{2} \rmi \gamma^{a} E_{a} \e^i + \tfrac{1}{2} N \e^{i}  - \g \cdot T L^{ij} \e_{j} + 3 L^{ij}\eta_{j}\,,\nn\\
\delta E_{a} &=& -\tfrac{1}{2} \rmi \bar{\e} \g_{ab} \mathcal{D}^{b} \varphi  - 2 \bar{\e} \gamma^{b} \varphi T_{ba} - 2\bar{\eta} \g_{a} \varphi\,, \nn\\
\delta N &=& \tfrac{1}{2} \bar{\e} \slashed{\mathcal{D}}\varphi  + \tfrac{3}{2} \rmi \bar{\e} \g \cdot T \vf + 4 \rmi \bar{\e}^i \chi^j L_{ij} + \tfrac{3}{2} \rmi \bar{\eta} \varphi\,,
\label{trlm}
\end{eqnarray}
where the following superconformally covariant derivatives are used
\bea
\mathcal{D}_\m L^{ij}&=&(\partial_{\m}-3b_{\m})L^{ij}+2V_{\m}{}^{(i}{}_kL^{j)k}-\rmi\bar{\p}_{\m}^{(i}\varphi^{j)}\,, \nn \\
\mathcal{D}_\m \varphi^i&=&(\partial_{\m}-\tfrac{7}{2}b_\m +\ft14\o_\m{}^{ab}\g_{ab})\varphi^i-V_{\m}^{ij}\varphi_j+\tfrac{1}{2} \rmi \slashed{\mathcal{D}}L^{ij}\p_{\m\,j} + \tfrac{1}{2} \rmi \gamma^{a} E_{a} \p_\m^i \nn \\
&&- \tfrac{1}{2} N \p_\m^{i}  + \g \cdot T L^{ij} \p_{\m\,j} - 3 L^{ij}\phi_{\m\,j}\,, \nn \\
\mathcal{D}_\m E_a &=&(\partial_\m-4b_\m)E_a+\o_{\m ab}E^b+\tfrac{1}{2} \rmi \bar{\p}_\m \g_{ab} \mathcal{D}^{b} \varphi  +2 \bar{\p}_\m \gamma^{b} \varphi T_{ba} + 2\bar{\phi}_\m \g_{a} \varphi\,.
\eea
As mentioned before, $E_a$ is constrained for the closure of the superconformal algebra
\begin{eqnarray}
\mathcal{D}^{a} E_{a}= 0\,. \label{closure}
\end{eqnarray}
Therefore $E_a$ can be solved in terms of a 3-form $E_{\m\n\r}$ as
\bea
E^a = - \tfrac{1}{12}e_{\m}{}^{a}e^{-1}\varepsilon^{\m\n\r\s\l}\mathcal{D}_\n  E_{\r\s\l}.\,
\label{3formE}
\eea
Similar to the vector multiplet, if the dilaton Weyl multiplet is adopted, the supersymmetry transformation rules for the linear multiplet can be obtained by using the map (\ref{UMap}).


\section{Superconformal Actions}\label{section: sactions}

In this section, we review the superconformal actions for the matter multiplets coupled to the standard Weyl multiplet and the dilaton Weyl multiplet, which give rise to different formulations of off-shell Poincar\'e supergravities.  These actions include a linear multiplet action \cite{Coomans:2012cf, Ozkan:2013uk} and three vector multiplet actions. The first vector multiplet action describing $n$ number of vector multiplets coupled to the standard Weyl is treated as the master action from which we derive the other two actions for $n$ number of vector multiplets coupled to the dilaton Weyl multiplet.

 The starting point for the constructions of the linear and vector multiplet actions is the superconformal density formula \cite{Fujita:2001kv}
\bea
e^{-1} \cL_{VL} &=& Y^{ij} L_{ij} + \rmi \bar\l \vf - \frac12 \bar\p_a^i \g^a \l^i L_{ij} + A_a P^a \nn\\
&& + \rho \Big(N + \frac12 \bar\p_a \g^a \vf + \frac14 \rmi \bar\p_a^i \g^{ab} \p_b^j L_{ij}\Big)\,,
\label{densityformula}
\eea
where $P_a$ is the bosonic part of the supercovariant field strength $E_a$
\bea
P^a &=& E^a + \frac12 \rmi \bar\p_b \g^{ba} \vf + \frac14 \bar\p_b^i \g^{abc} \p_c^j L_{ij}.
\label{PmEm}
\eea


\subsection{The Linear Multiplet Action}\label{ss:laction}

The procedure of constructing an action for the linear multiplet is based on superconformal tensor calculus, where the fields of vector multiplets are composed in terms of the fields in linear multiplet and a Weyl multiplet. The construction of a linear multiplet action is explored in detail by \cite{Coomans:2012cf, Ozkan:2013uk}. Here we only present the results which contribute to the purely bosonic action. Using the linear multiplet and the standard Weyl multiplet, the elements of vector multiplet can be composed as given in (\ref{LEmbde}). Inserting the composite expressions into the vector-linear action (\ref{densityformula}), one can obtain an action for the linear multiplet,
\bea
e^{-1}{\cal{L}}_L^S &=& L^{-1} L_{ij} \Box^c L^{ij} - L^{ij} {\cal{D}}_{\m}L_{k(i} {\cal{D}}^{\m} L_{j)m} L^{km} L^{-3} + N^2 L^{-1}  \nn\\
&& - E_{\m} E^{\m} L^{-1} + \tfrac{8}{3} L T^{2} + 4 D L - \tfrac{1}{2}L^{-3} E^{\m\n} L_{k}^{l} \partial_{\m} L^{kp} \partial_\n L_{pl}  \nn\\
&& + 2 E^{\m\n} \partial_{\m} ( L^{-1} E_{\n} + V_{\n}^{ij} L_{ij} L^{-1} )\,,
\label{Lact}
\eea
where we have defined $L^2 = L_{ij} L^{ij}$ and the superscript $S$ indicates that this action utilizes the standard Weyl multiplet. The superconformal d'Alembertian of $L^{ij}$ is given by
\bea
L_{ij} \Box^c L^{ij}&=&L_{ij}(\partial^a-4b^a+\omega_b{}^{ba}){\cal{D}}_a L^{ij}+2L_{ij} V_a{}^i{}_k{\cal{D}}^a L^{jk}+6L^2f_a{}^a \nn \\
&&-iL_{ij}\bar{\psi}^{a i}{\cal D}_a\varphi^j-6L^2\bar{\psi}^a\g_a\chi-L_{ij}\bar{\varphi}^i\g\cdot T\g^a\psi_a^j+L_{ij}\bar{\varphi}^i\g^a\phi_a^j\,.
\eea

The linear multiplet action (\ref{Lact}) can be transferred to an action describing the linear multiplet coupled to the dilaton Weyl multiplet by replacing the $T_{ab}, D$ and $\chi^i$ by their composite expressions according to (\ref{UMap}) in the action (\ref{Lact}).


\subsection{Vector Multiplet Actions}\label{ss:vactions}

The elements of linear multiplet can be constructed in terms of the fields in vector multiplet and a Weyl multiplet (\ref{VEmbed}). Then an action for vector multiplet can be obtained by using the density formula (\ref{densityformula}). Adopting the standard Weyl multiplet, an action for $n$ vector multiplets is given by \cite{Bergshoeff:2002qk, Fujita:2001kv}
\bea
e^{-1} \cL_{V}^S &=& C_{IJK} \Big( - \ft14 \r^I F_{\m\n}^J F^{K\,\m\n} + \ft13 \r^I \r^J \Box^C \r^K +\ft16 \r^I \cD_\m \r^J \cD^\m \r^K + \r^I Y^{J\, ij} Y_{ij}^K\nn\\
&& \qquad \qquad - \ft43 \r^I \r^J \r^K (D + \ft{26}3 T_{\m\n} T^{\m\n}) + 4 \r^I \r^J F_{\m\n}^K T^{\m\n} - \ft1{24} \e^{\m\n\r\s\l} A_{\m}^I F_{\n\r}^J F_{\s\l}^K \Big),
\label{Vact}
\eea
where we have generalized the single vector multiplet action to $n$-vector multiplets action, $I = 1, \ldots n$. The coefficient $C_{IJK}$ is symmetric in $I,J,K$  and determines the coupling of $n$ vector multiplets. The complete expression of the superconformal d'Alembertian for $\r^I$ is \cite{Bergshoeff:2002qk}
\bea
\Box^c \r^I &=& (\partial^a - 2b^a + \o_b{}^{ba}) \cD_a \r^I - \ft12 \rmi \bar\p_a \cD^a \l^I - 2\r^I \bar\p_a \g^a {\chi} \nn\\
&& + \ft12 \bar\p_a \g^a \g \cdot {T} \l^I + \ft12 \bar\phi_a \g^a \l^I + 2 f_a{}^a \r^I.
\eea
The action (\ref{Vact}) will be used as the \emph{master action} for the construction of $n$-vector coupled curvature squared actions in the dilaton Weyl multiplet. Adopting the dilaton Weyl multiplet, the fields $D, T_{ab}$ and $\chi^i$ are underlined meaning that they become composite fields, one obtains a vector multiplet action describing $n$ number of vector multiplets coupled to supergravity. As we will see, the vector multiplet action plays an important role in the construction of Riemann squared invariant \cite{Bergshoeff:2011xn,Ozkan:2013uk}. A special case of the \emph{master action} to be utilized directly in the derivation of Riemann squared invariant is given by
\bea \label{CIJKChoice}
C_{IJK} = \left\{ \begin{array}{lll}
C_{1IJ} &  = &  a_{IJ}   \\
 0 & & \textrm{otherwise}. \\
\end{array} \right.
\eea
Under this choice, the vector multiplet action in the dilaton Weyl multiplet is given as
\bea
e^{-1} \cL_{V}^{'D} &=& a_{IJ} \Big( \r Y_{ij}^I Y^{ij\, J} - \ft14 \r {F}_{\m\n}^I {F}^{\m\n\, J} - \ft12 \r^I  {F}^J_{\m\n} {F}^{\m\n} + 8 \r \r^I {F}^J_{\m\n} \underline{T}^{\m\n} + \ft12 \r^I \r^J \Box^C \r \nn\\
&& \quad   + \ft12 \r \r^I \Box^C \r^J + \ft12 \r^I \cD_a \r^J \cD^a \r - 4 \r \r^I \r^J (\underline{D} + \ft{26}3 \underline{T}^2 ) + 4  \r^I \r^J {F}_{\m\n}\underline{ T}^{\m\n} \nn\\
&& \quad - \ft18  \e^{\m\n\r\s\l} F_{\m\n}^I F_{\r\s}^J A_\l + 2 \r^I Y^J_{ij} Y^{ij} \Big)\,,
\label{NVDW}
\eea
where the superscript $D$ demonstrates that this action is constructed by using the dilaton Weyl multiplet. In this case, the coupling of $n$ number of vector multiplets is dictated by the symmetric rank-2 tensor $a_{IJ}$. To obtain the Riemann squared invariant by using the Yang-Mills trick (\ref{YMDWMap}), the $I$ index indicating $n$-copy of Abelian vector multiplets should be replaced by the Yang-Mills index $\Sigma$ and the field strength should be interpreted as the Yang-Mills field strength . As we shall mention later, this vector multiplet action can give rise to the Riemann squared invariant coupled to $n$ number of vector multiplets. If one prefers to construct a Riemann squared action purely based on the dilaton Weyl multiplet \cite{Bergshoeff:2011xn,Ozkan:2013uk} one can utilize the map from the vector multiplet to the dilaton Weyl multiplet \cite{Ozkan:2013uk}
 \be
(\l_i,\r,A_\m,Y_{ij})\rightarrow(\p_i,\s,C_\m,\ft14 \rmi \s^{-1} \bar\p_i \p_j ).
\label{DVMap}
\ee
which leads to another action for the vector multiplet coupled to supergravity
\bea
e^{-1} \cL_{V}^{D} &=& a_{IJ} \Big( \s Y_{ij}^I Y^{ij\, J} - \ft14 \s {F}_{\m\n}^I {F}^{\m\n\, J} - \ft12 \r^I  {F}^J_{\m\n} {G}^{\m\n} + 8 \s\r^I {F}^J_{\m\n} T^{\m\n} + \ft12 \r^I \r^J \Box^C \s \nn\\
&& \quad   + \ft12 \s \r^I \Box^C \r^J + \ft12 \r^I \cD_{\m} \r^J \cD^{\m} \s - 4 \s \r^I \r^J (\underline{D} + \ft{26}3 \underline{T}^2 ) + 4  \r^I \r^J {G}_{\m\n}\underline{ T}^{\m\n} \nn\\
&& \quad - \ft18  \e^{\m\n\r\s\l} F_{\m\n}^I F_{\r\s}^J C_\l \Big)\,.
\label{VDW}
\eea


\section{Minimal Curvature Squared Actions in the Dilaton Weyl Multiplet}\label{section: alldilaton}
The five dimensional minimal off-shell Poincar\'e supergravity multiplet consists of the fields
\be
e_\m{}^a(10),\, \p_\m^i(32),\, C_\m(4),\, B_{\m\n}(6),\,\vf^i(8),\,L(1),\, E_{\m\n\r}(4) ,\, N(1),\, V_\m(4) ,\, V_\m^{'ij}(10),
\label{fieldcontent}
\ee
where the number in the bracket denotes the off-shell degrees of freedom carried by the fields.
Using the dilaton Weyl multiplet, the minimal off-shell supersymmetric Riemann squared and Gauss-Bonnet combination have been obtained in \cite{Bergshoeff:2011xn, Ozkan:2013uk} and \cite{Ozkan:2013uk} respectively. In this section, we complete the off-shell curvature squared invariants by constructing the Ricci scalar squared action. The map from the dilaton Weyl multiplet to the standard Weyl multiplet (\ref{UMap}) plays a crucial role in the construction of curvature squared actions. In particular, the composite expression for $D$ contains a curvature term. Thus, the existence of a $D^2$ term in a curvature squared action means the curvature terms get an extra $R^2$ contribution from the composite expression of $D$. As we shall see, this fact is essential in the construction of supersymmetric completion of Gauss-Bonnet combination \cite{Ozkan:2013uk}.

In the first two subsections, we list the known supersymmetric Riemann squared and Gauss-Bonnet actions. In the third subsection, we obtain the supersymmetric Ricci scalar squared action from superconformal method.


\subsection{Supersymmetric Riemann Squared Action}

In this subsection, we briefly revisit the Riemann squared action constructed in \cite{Ozkan:2013uk} using superconformal calculus and Yang-Mills trick in five dimensions
\bea
( A_\m^{\Sigma}, \quad Y_{\Sigma}^{ij}, \quad \l_{\Sigma}^i, \quad \r_{\Sigma}) \quad \longleftrightarrow \quad ( \o_{\m +}^{ab}, \quad - \widehat{V}_{ab}{}^{ij}, \quad - \widehat\p_{ab}^i, \quad \widehat{G}_{ab} ),
\label{YMDWMap}
\eea
where $\Sigma$ means the Yang-Mills index.
This action can also be obtained from a circle reduction of six
dimensional theory \cite{Bergshoeff:2011xn}. The derivation of the Riemann squared action requires specific gauge fixing conditions given by
\bea
\s = 1, \qquad \p^i = 0, \qquad L_{ij} = \ft1{\sqrt2} \d_{ij}L, \qquad b_\m = 0\,.
\label{GF2}
\eea
The first gauge choice breaks the $\SU(2)_R$ down to $\U(1)_R$ whereas the second one fixes
dilatations, the third one fixes special supersymmetry transformations and the last one
fixes conformal boosts. The decomposition rules which follow from the above gauge fixing can be found in \cite{Ozkan:2013uk}. As a consequence of the gauge fixing, we obtain the Poincar\'e multiplet. Accordingly, the off-shell supersymmmetry transformation rules for the fields of Poincar\'e multiplet are given by \cite{Ozkan:2013uk}
\bea
\d e_{\m}{}^a &=& \ft12 \bar\e \g^a \p_\m \, ,\nn\\
\d \p_\m^i &=& \cD_\m (\o_{-}) \e^i  - \ft12 \rmi \widehat{G}_{\m\n} \g^\n \e^i \,, \nn\\
\d V_{\m}{}^{ij} &=& \ft12 \bar\e^{(i} \g^\n \widehat\p_{\m\n}^{j)} - \ft16 \bar\e^{(i} \g \cdot \widehat{H} \p_\m^{j)} - \ft14 \rmi \bar\e^{(i} \g \cdot \widehat{G} \p_\m^{j)},  \nn\\
\d C_\m &=& -\ft12\rmi\bar\e \p_\m \, ,\nn\\
\d B_{\m\n} &=& \ft12 \bar\e \g_{[\m} \p_{\n]} + C_{[\m} \d(\e) C_{\n]} \,,\nn\\
\d L &=& \ft1{\sqrt2} \rmi \bar\e^i \vf^j \d_{ij} \,,\nn\\
\d \vf^i &=& - \ft1{2\sqrt2} \rmi \slashed{\partial} L \d^{ij} \e_j - \ft1{\sqrt2} \rmi V'_{\m}{}^{(i}{}_k \d^{j)k} L \e_j  - \ft12 \rmi \slashed{E} \e^i  + \ft12 N \e^i + \ft1{4\sqrt2} L \g \cdot \widehat{G} \d^{ij} \e_j \, \nn\\
&&  - \ft1{6\sqrt2} \rmi L \g \cdot \widehat{H} \d^{ij} \e_j  ,\nn\\
\d E_{\m\n\r} &=& - \bar\e \g_{\m\n\r} \vf + \ft1{\sqrt2} \rmi L \bar\p^i_{[\m} \g_{\n\r ]} \e^j \d_{ij} \,, \nn\\
\d N &=& \ft12 \bar\e \g^\m \Big( \partial_\m + \ft14 \o_\m{}^{bc} \g_{bc} \Big) \vf + \ft12 \bar\e^i \g^a V_{a\,ij} \vf^j - \ft1{4\sqrt2} \rmi \bar\e^i \g^a \slashed{\partial}L  \p_a^j \d_{ij}\nn\\
&& + \ft1{4\sqrt2}\rmi \bar\e^i \g^a \g^b V'_{b(i}{}^k \d_{j)k} \p_a^j + \ft14 \rmi \bar\e \g^a \slashed{E} \p_{a} - \ft14 N \bar\e  \g^a \p_a + \ft1{8\sqrt2} L \bar\e^i \g^a \g \cdot \widehat{G} \p_a^j \d_{ij}\nn\\
&&-\sqrt2 L \bar\e^i \g^a \phi_a^j \d_{ij} + \ft18 \rmi \bar\e \g \cdot \widehat{H} \vf ,
\label{UnGaugedTransform}
\eea
The bosonic part of the Riemann squared action is given as \cite{Ozkan:2013uk}
\bea
e^{-1}{\cal L}_{{\rm Riem}^2}^{D}  &=& -\ft14
\Big(\,R_{\mu\nu ab}(\omega_+)- G_{\mu\nu} G_{ab}\Big)
\left(\,R^{\mu\nu ab}(\omega_+)- G^{\mu\nu} G^{ab}\right) \nn\\ &&
-\ft12 \nabla_\mu(\omega_+) G^{ab} \nabla^\mu(\omega_+)G_{ab}
+V_{\mu\nu}{}^{ij} V^{\mu\nu}{}_{ij} \nn\\ && - \ft1{8}
\e^{\mu\nu\rho\sigma\lambda} \Big(\,R_{\mu\nu
ab}(\omega_+)- G_{\mu\nu} G_{ab}\Big)
\left(\,R_{\rho\sigma}{}^{ab}(\omega_+)- G_{\rho\sigma}
G^{ab}\right) C_\lambda \nn\\ && - \ft12
\e^{\mu\nu\rho\sigma\lambda} B_{\rho\sigma}\left(\,R_{\mu\nu
ab}(\omega_+)- G_{\mu\nu}G_{ab}\right) \nabla_\lambda (\omega_+)
G^{ab}\,,
\label{bosonicR2}
 \eea
where the torsionful spin connection \cite{Bergshoeff:2011xn} is defined as
\bea
{\omega}_\mu{}^{ab}_\pm &=& {\omega}_\mu{}^{ab} \pm {\widehat
H}_\mu{}^{ab}\ .
\label{torsion} \eea
We have also defined the supercovariant curvature of $V_\m{}^{ij}$ as
\bea {\widehat V}_{\mu\nu}{}^{ij} &=& V_{\mu\nu}{}^{ij} -
{\bar\psi}_{[\mu}^{(i}\gamma^\rho {\widehat\psi}_{\nu]\rho}^{j)} +\ft1{6} {\bar\psi}_\mu^{(i} \gamma\cdot {\widehat
H}\psi_\nu^{j)} +\ft14 i {\bar\psi}_\mu^{(i} \gamma\cdot {\widehat
G} \psi_\nu^{j)}\ . \label{vmn}
\eea
The Riemann squared action can be added to the off-shell Poincar\'e supergravity that is also invariant under the transformation rules (\ref{UnGaugedTransform}) \cite{Ozkan:2013uk}
\bea
e^{-1}{\cL}_{LR}^D	 &=& \tfrac12 L R + \ft12 L^{-1} \partial_\m L \partial^\m L- \ft14 L G_{\m\n} G^{\m\n} - \ft{1}6 L H_{\m\n\r} H^{\m\n\r}\nn\\
&& - L^{-1} N^2  - L^{-1} P_\m P^\m - \sqrt2 P^\m V_\m +  L V_\m^{'ij} V^{'\m}_{ij}\,,
\label{RLag}
\eea
where $P_\m$ is the bosonic part of the supercovariant curvature $E_\m$ according to (\ref{PmEm}), and we have decomposed the field $V_\m^{ij}$ into its trace and traceless part as
\bea
V_\m^{ij} = V_\m^{'ij} + \tfrac12 \d^{ij} V_\m, \qquad  V_\m^{'ij} \d_{ij}=0\,.
\eea

\subsection{Supersymmetric $C^2_{\m\n\r\l}+\ft16R^2$ Action}

In this section, we review the supersymmetrization of Weyl tensor squared action in the dilaton Weyl multiplet. As we mentioned before, in the dilaton Weyl multiplet, the supersymmetrization of Weyl tensor squared accquires an extra Ricci scalar squared term through the square of $D$. Therefore, the supersymmetric completion of Weyl tensor squared $C_{\m\n\r\s}^2$ turns into the supersymmetric completion of $C^2_{\m\n\r\l}+\ft16R^2$. The bosonic part of the supersymmetric $C^2_{\m\n\r\l}+\ft16R^2$ action is presented by \cite{Ozkan:2013uk}
\bea \label{pregb2}
e^{-1} \cL_{C^2 + \ft16R^2}^D &=& \ft18  R_{\m\n\r\s} R^{\m\n\r\s} - \ft16  R_{\m\n} R^{\m\n} + \ft1{48}  R^2 + \ft{64}3 \underline{D}^2 + \ft{1024}9  \underline{T}^2 \underline{D}\nn\\
&&  - \ft{16}3  {R}_{\m\n\r\s} \underline{T}^{\m\n} \, \underline{T}^{\r\s} + 2{R}_{\m\n\r\s} \underline{T}^{\r\s} G^{\m\n} + \ft13 R \underline{T}_{\m\n} G^{\m\n} - \ft83 R_{\n\s} G_\rho{}^\nu \underline{T}^{\r\s} \nn\\
&&  - \ft{64}3 R^{\n\r} \underline{T}_{\m\n} \underline{T}^\m{}_\r  +\ft{8}3 R \underline{T}^2  - \ft{32}3 \underline{D} \, \underline{T}_{\m\n} G^{\m\n}  + \ft1{16} \e_{\m\n\r\s\l}C^\m {R}^{\n\r\t\d} {R}^{\s\l}{}_{\t\d} \nn\\
&& -\ft1{12} \e_{\m\n\r\s\l} C^\m V^{\n\r}{}_{ij} V^{\s\l\, ij}  - \ft{1}3 V_{\m\n}{}^{ij} V^{\m\n}{}_{ij} -\ft{64}3   \nabla_\m \underline{T}_{\n\r} \nabla^\m \underline{T}^{\n\r} \nn\\
&&+ \ft{64}3  \nabla_\n \underline{T}_{\m\r} \nabla^\m \underline{T}^{\n\r} - \ft{128}3  \underline{T}_{\m\n} \nabla^\n \nabla_\r \underline{T}^{\m\r} - \ft{128}3 \e_{\m\n\r\s\l} \underline{T}^{\m\n} \underline{T}^{\r\s} \nabla_\t \underline{T}^{\l\t} \nn\\
&&+ 1024  \, \underline{T}^4- \ft{2816}{27}  (\underline{T}^2)^2  - \ft{64}9 \underline{T}_{\m\n} G^{\m\n} \underline{T}^2 - \ft{256}3 \underline{T}_{\m\r} \underline{T}^{\r\s} \underline{T}_{\n\s} G^{\m\n}   \nn\\
&&  - \ft{32}3   \e_{\m\n\r\s\l}  \underline{T}^{\r\t} \nabla_\t \underline{T}^{\s\l} G^{\m\n}- 16   \e_{\m\n\r\s\l} \underline{T}^\rho{}_\t \nabla^\s \underline{T}^{\l\t} G^{\m\n}\,,
\eea
where the bosonic parts of the composite fields $\underline{D}$ and $\underline{T}^{ab}$ are given as
\bea
\underline{D}&\equiv&-\ft1{32}R-\ft1{16}G^{ab}G_{ab}-\ft{26}3 \underline{T}^{ab}\underline{T}_{ab}+2\underline{T}^{ab}G_{ab}\,,\cr
\underline{T}_{ab} &\equiv& \ft18 G_{ab} + \ft1{48} \e_{abcde} H^{cde}\,.
\label{UMap2}
\eea
For simplicity, we introduced the notation $T^4 \equiv T_{ab} T^{bc} T_{cd} T^{da}$ and $(T^2)^2 \equiv (T_{ab} T^{ab})^2$. We also used
\bea
\nabla_\m T_{ab} = \partial_\m T_{ab} - 2 \o_\m{}^c{}_{[a} T_{b]c}.
\eea
Note that to obtain the above expressions for the composite fields, the gauge fixing conditions (\ref{GF2}) have been utilized. The supersymmetric completion of $C_{\m\n\r\s}^2 + \ft16 R^2$, (\ref{pregb2}) can be combined with the Riemann squared action (\ref{bosonicR2}) and the Poincar\'e supergravity (\ref{RLag})
\bea
\cL_{LR}^{D} + \alpha \cL_{{\rm Riem}^2}^D + \beta \cL_{C^2+\ft16R^2}^D.
\eea
This theory possesses a maximally supersymmetric ${\rm Minkowski}_5$ vacuum. The spectrum around the ${\rm Minkowski}_5$ vacuum has been analyzed in \cite{Ozkan:2013uk}, where it is found that for generic $\alpha,\,\beta$, the full spectrum consists of the (reducible) massless 12+12 supergravity multiplet with fields
$(h_{\mu\nu},\,b_{\mu\nu},\,c_{\mu},\,\phi,$ $\,\psi^{i}_{\mu},\,\varphi^i)$ and a ghost massive 32+32 supergravity
multiplet with fields $(h_{\mu\nu},\,b_{\mu\nu},\,c_{\mu},\,\phi,\,v^{ij}_{\mu},\psi^{i}_{\mu},\,\varphi^i)$. In the special case when $\b = 3 \a$, the massive multiplet decouples and the curvature squared terms furnish the supersymmetric completion of Gauss-Bonnet combination \cite{Ozkan:2013uk}.

\subsection{Supersymmetric Ricci Scalar Squared Action}

In this section, we construct the supersymmetric completion of Ricci scalar squared action using the dilaton Weyl multiplet. The key observation behind the construction is that the composite expression of $Y^{ij}$ (\ref{LEmbde}) contains the Ricci scalar implicitly in the superconformal d'Alembertian of $L^{ij}$. Therefore, the supersymmetric Ricci scalar squared action can be obtained by substituting the composite expressions (\ref{LEmbde}) in the vector multiplet action given in (\ref{VDW}) since the off-shell vector multiplet action has a $Y^{ij}Y_{ij}$ term. The construction of supersymmetric Ricci scalar squared action completes the off-shell curvature squared actions based on the dilaton Weyl multiplet in $\cN=2,\, D=5$ supergravity.

For the construction of Ricci scalar squared invariant, we first rewrite the vector multiplet action (\ref{VDW}) for a single vector multiplet under the gauge fixing conditions (\ref{GF2}) as,
\bea
e^{-1} \cL_V^D|_{\s=1} &=& Y_{ij} Y^{ij} - \ft12 \nabla_\m \r \nabla^\m \r - \ft14 (F_{\m\n} - \r G_{\m\n}) (F^{\m\n} - \r G^{\m\n}) \nn\\
&& - \ft18 \e^{\m\n\r\s\l} (F_{\m\n} - \r G_{\m\n})(F_{\r\s} - \r G_{\r\s}) C_\l \nn\\
&& - \ft12 \e^{\m\n\r\s\l} (F_{\m\n} - \r G_{\m\n}) B_{\r\s} \nabla_\l \rho \,.
\label{fixedVDW}
\eea
Using the same gauge fixing, the composite expressions (\ref{LEmbde}) for the elements of vector multiplet can be recast into
\bea
\underline{\r}|_{\s=1} &=& 2 N L^{-1},  \nn\\
\underline{Y}_{ij}|_{\s=1} &=& \ft1{\sqrt2} \d_{ij} \Big(- \ft12 R + \ft14 G_{ab} G^{ab} + \ft16 H_{abc} H^{abc} - L^{-2} N^2 - L^{-2} P_{a}P^{a} - V_{a}^{'kl} V^{'a}_{kl} \nn\\
&& \qquad  + L^{-1} \Box L - \ft12 L^{-2} \partial_a L \partial^a L \Big)  + 2L^{-1} P^a V'_{aij}  - \sqrt{2} L^{-1}\nabla^a(L V'_{a}{}^m{}_{(i} \d_{j)m}), \nn\\
\underline{\widehat{F}}_{ab}|_{\s=1} &=& 2\sqrt{2} \partial_{[a} \Big( V_{b]} + \sqrt2 L^{-1} P_{b]} \Big).
\eea
The fermionic terms in the composite expressions of vector multiplet can be straightforwardly figured out by using the complete results given in (\ref{LEmbde}). Using the above formulas in (\ref{fixedVDW}), we obtain the supersymmetric Ricci scalar squared action in the dilaton Weyl multiplet whose bosonic part reads
\bea
e^{-1} \cL_{R^2}^{D} &=& \ft14 \Big( R - \ft12 G_{\m\n} G^{\m\n} - \ft13 H_{\m\n\r} H^{\m\n\r} + 2 L^{-2} N^2 + 2 L^{-2} P_\m P^\m -4Z_{\m}\bar{Z}^{\m}- 2 L^{-1} \Box L \nn\\
&& + L^{-2} \partial_\m L \partial^\m L \Big)^2  -L^{-2} \Big|2\nabla^{\m}(L Z_{\m})+ 2\sqrt{2}\rmi P^{\m}Z_{\m} \Big|^2 - 2 \nabla_\m(L^{-1}N) \nabla^\m(L^{-1}N) \nn\\
&& -\ft12 \e^{\m\n\r\s\l} \Big( \partial_{\m} \widetilde{C}_{\n} - NL^{-1} G_{\m\n} \Big)\Big( \partial_{\r} \widetilde{C}_{\s} - NL^{-1} G_{\r\s}  \Big) C_\l \nn\\
&& - 2\e^{\m\n\r\s\l}\Big(\partial_{\m} \widetilde{C}_{\n} - NL^{-1} G_{\m\n} \Big)B_{\r\s} \nabla_\l (L^{-1} N)\nn\\
&& - \Big( \partial_{[\m} \widetilde{C}_{\n]} - NL^{-1} G_{\m\n} \Big)\Big( \partial^{\m} \widetilde{C}^{\n} - NL^{-1} G^{\m\n} \Big), \,
\eea
where for simplicity, we have defined
\be
Z_{\m}=V^{'12}_{\m} + \rmi V^{'11}_{\m},\qquad \widetilde{C}_{\mu} =\sqrt{2}V_{\m}+2L^{-1}P_{\m}.
\ee
The general $R+R^2$ action in the dilaton Weyl multiplet can therefore be written as
\bea
\cL_{LR}^{D} + \alpha \cL_{{\rm Riem}^2}^D + \beta \cL_{C^2+\ft16R^2}^D+\gamma \cL_{R^2}^D.
\eea
The inclusion of the Ricci scalar squared action does not affect the existence of maximally supersymmetric ${\rm Minkowski}_5$ vacuum, however it brings a massive vector multiplet with $m^2=\frac{L_0}{2\gamma}$. The 8+8 degrees of freedom in the massive vector multiplet are carried by $(L,N,\partial^{\m}Z_{\m},\widetilde{C}_{\m},\varphi^i)$.

\section{Vector Multiplets Coupled Curvature Squared Invariants in the Dilaton  Weyl Multiplet}\label{section: dilationvector}
In the previous section, we have completed the curvature squared invariants purely based on the off-shell Poincar\'e multiplet (\ref{fieldcontent}). In this section, we couple the external vector multiplets to the curvature squared invariants. The inclusion of the external vector multiplet gives rise to a mixed Chern-Simons term in the supersymmetric Riemann squared action
\bea
A \wedge R \wedge R \,,
\eea
where the vector $A_{\m}$ belongs to a vector multiplet, as opposed to the case of minimal off-shell curvature squared invariants in the dilaton Weyl multiplet where the Chern-Simons term is purely gravitational (\ref{bosonicR2})
\bea
C \wedge R \wedge R \,,
\eea
where $C_{\m}$ is the vector in the Poincar\'e multiplet. In the following, we directly present the results for the vector multiplets coupled curvature squared term which can be straightforwardly obtained from the results for single vector multiplet coupled curvature squared term.
\subsection{Vector Multiplets Coupled Riemann Squared Action}
In this subsection, we generalize the Riemann squared action purely based on the off-shell Poincar\'e multiplet to the vector multiplets coupled Riemann squared action in which the Chern-Simons term takes the form of $A\wedge R \wedge R$. In order to construct the vector multiplets coupled Riemann squared action, we consider the following Yang-Mills action in the dilaton Weyl multiplet. This action is the Yang-Mills analogue of the $n$ Abelian vector action (\ref{NVDW})
\bea
e^{-1} \cL_{\rm YM}^{'D} &=&  \r Y_{ij}^\S Y^{ij\,\S} + 2\r^\S Y_{ij}^\S Y^{ij} + \r \r^\S \nabla_\m \nabla^\m \r^\S  + \ft12 \r \nabla_\m \r^\S \nabla^\m \r^\S  \nn\\
&& -\ft14 \r ( F_{\m\n}^\S - \r^\S G_{\m\n}) ( F^{\m\n\,\S} - 3 \r^\S G^{\m\n}) -\ft12 (F_{\m\n}^\S -  \r^\S G_{\m\n} )\r^\S F^{\m\n} \nn\\
&& + \ft1{12}  \r^\S \r^\S \e^{\m\n\r\s\l}(F_{\m\n} - 2\r G_{\m\n}) H_{\r\s\l} + \ft16 \r \r^\S \e^{\m\n\r\s\l} F_{\m\n}^\S H_{\r\s\l} \nn\\
&& -\ft18 \e^{\m\n\r\s\l} F_{\m\n}^\S F_{\r\s}^\S A_\l \,,
\label{NVS1}
\eea
where $\S$ is the Yang-Mills group index. The construction procedure of the vector multiplets coupled Riemann squared action is the same as before. Upon applying the map between the Yang-Mills multiplet and the dilaton Weyl multiplet, we obtain the vector multiplets coupled Riemann squared action
\bea
&&e^{-1} \cL_{\rm Riem^2}^{'D}= \a_I \Big[-\ft14 \r^I ( R_{\m\n ab}(\o_+) - G_{\m\n} G_{ab}) (R^{\m\n ab}(\o_+) - 3 G^{\m\n} G^{ab}) \nn\\
&& \qquad\qquad\quad - \ft12 (R^{\m\n ab}(\o_+) - G^{\m\n} G^{ab}) F^I_{\m\n} G_{ab} + \r^I V_{ij}{}^{\m\n} V^{ij}{}_{\m\n} - 2 G_{\m\n} V_{ij}{}^{\m\n} Y^{ij\,I}  \nn\\
&& \qquad\qquad\quad  + \r^I G_{ab} \nabla_\m(\o_+) \nabla^\m(\o_+) G^{ab}  + \ft12 \r^I \nabla_\m (\o_+) G^{ab} \nabla^\m (\o_+) G_{ab}  \nn\\
&& \qquad\qquad\quad  + \ft1{12} \e^{\m\n\r\s\l}(F_{\m\n}^I - 2\r^I G_{\m\n}) H_{\r\s\l} G_{ab} G^{ab} + \ft16 \r^I \e^{\m\n\r\s\l} R_{\m\n ab}(\o_+) G^{ab}H_{\r\s\l} \nn\\
&& \qquad\qquad\quad  - \ft18 \e^{\m\n\r\s\l} R_{\m\n ab}(\o_+) R_{\r\s}{}^{ab}(\o_+) A_\l^I \Big].
\eea
Note that this action recovers the Riemann squared invariant (\ref{bosonicR2}) upon considering a single vector multiplet, $I=1$, and applying the map from the dilaton Weyl multiplet to the vector multiplet (\ref{DVMap}).

\subsection{Vector Multiplets Coupled $C^2_{\m\n\r\l}+\ft16R^2$ Action}
The $n$-vector multiplets coupled $C^2_{\m\n\r\l}+\ft16R^2$ action can be straightforwardly obtained from the Weyl squared action \cite{Hanaki:2006pj} in standard Weyl multiplet by underlining $D$, $T_{ab}$ and $\chi_i$
\bea
e^{-1} \cL_{C^2+\ft16 R^2}^{'D} &=&  \b_{I} \Big[ \ft18\r^I {C}^{\m\n\r\s} {C}_{\m\n\r\s}+ \ft{64}3 \r^I \underline{D}^2 + \ft{1024}9 \r^I \underline{T}^2 \underline{D} - \ft{32}3 \underline{D} \, \underline{T}_{\m\n} F^{\m\n\,I}   \nn\\
&&  - \ft{16}3 \r^I {C}_{\m\n\r\s} \underline{T}^{\m\n} \, \underline{T}^{\r\s} + 2{C}_{\m\n\r\s} \underline{T}^{\m\n} F^{\r\s\,I} + \ft1{16} \e^{\m\n\r\s\l}A_\m^I {C}_{\n\r\t\d} {C}_{\s\l}{}^{\t\d}    \nn\\
&& -\ft1{12} \e^{\m\n\r\s\l} A_\m^I {V}_{\n\r}{}^{ij} {V}_{\s\l\, ij} +  \ft{16}3 Y^I_{ij} {V}_{\m\n}{}^ {ij} \underline{T}^{\m\n} - \ft{1}3 \r^I {V}_{\m\n}{}^{ij}{V}^{\m\n}{}_{ij} \nn\\
&& +\ft{64}3 \r^I  \nabla_\n \underline{T}_{\m\r} \nabla^\m \underline{T}^{\n\r} - \ft{128}3 \r^I \underline{T}_{\m\n} \nabla^\n \nabla_\r \underline{T}^{\m\r} - \ft{256}9 \r^I R^{\n\r} \underline{T}_{\m\n} \underline{T}^\m{}_\r \nn\\
&& + \ft{32}9 \r^I R \underline{T}^2 - \ft{64}3  \r^I \nabla_\m \underline{T}_{\n\r} \nabla^\m \underline{T}^{\n\r} + 1024 \r^I \, \underline{T}^4- \ft{2816}{27} \r^I  (\underline{T}^2)^2   \nn\\
&&- \ft{64}9 {\underline{T}_{\m\n}} F^{\m\n\,I} \underline{T}^2 - \ft{256}3 \underline{T}_{\m\r} \underline{T}^{\r\l} \underline{T}_{\n\l} F^{\m\n\,I}  - \ft{32}3   \e_{\m\n\r\s\l}  \underline{T}^{\r\t} \nabla_\t \underline{T}^{\s\l} F^{\m\n\,I}  \nn\\
&& - 16   \e_{\m\n\r\s\l} \underline{T}^\rho{}_\t \nabla^\s \underline{T}^{\l\t} F^{\m\n\,I}  - \ft{128}3 \r^I \e_{\m\n\r\s\l} \underline{T}^{\m\n} \underline{T}^{\r\s} \nabla_\t \underline{T}^{\l\t}\Big] \,.
\label{pregb3}
\eea
where the composite expressions for $D$ and $T_{ab}$ in $\s=1$ gauge fixing are given in (\ref{UMap2}).
\subsection{Vector Multiplets Coupled Ricci Scalar Squared Action }
 To obtain an off-shell Ricci scalar squared invariant coupled to vector multiplets, we use the same strategy as we construct the minimal Ricci scalar squared invariant. The starting point is the vector multiplet action (\ref{NVDW}). By choosing the nonvanishing components of $C_{IJK}$ to be $C_{I11} = \a_I$  and replacing $D$, $T_{ab}$ by their composite expressions (\ref{UMap}), we obtain the $n$ vector coupled Ricci scalar squared action
\bea
e^{-1} \cL_{R^2}^{' D} &=& \g_I \Big( \r^I \underline{Y}_{ij} \underline{Y}^{ij} + 2 \underline{\r} \underline{Y}^{ij} Y_{ij}^I - \ft14 \r^I  \underline{F}_{\m\n}  \underline{F}^{\m\n} - \ft12 \underline{\r} \, \underline{F}^{\m\n} (F_{\m\n}^I - 2 \r^I G_{\m\n} ) \nn\\
&& \quad  + \ft12 \underline{\r}^2 ( F_{\m\n}^I - \ft32 \r^I G_{\m\n}) G^{\m\n} + \ft1{12}\underline{\r}^2 \e^{\m\n\r\s\l} (F_{\m\n}^I - 2 \r^I G_{\m\n} ) H_{\r\s\l} \nn\\
&& \quad + \ft16 \r^I \underline{\r} \e_{\m\n\r\s\l}\underline{F}^{\m\n} H^{\r\s\l} - \ft18 \e_{\m\n\r\s\l} A^{\m I} \underline{F}^{\n\r} \underline{F}^{\s\l} + \ft12 \r^I \nabla_\m \underline{\r} \nabla^\m \underline{\r}\nn\\
&& \quad + \r^I \underline{\r} \Box \underline{\r} \Big)\,.
\label{R2SW2}
\eea
The composite expressions for the elements of a vector multiplet are given as
 \bea
\underline{\r}|_{\s=1} &=& 2 N L^{-1},  \nn\\
\underline{Y}_{ij}|_{\s=1} &=& \ft1{\sqrt2} \d_{ij} \Big(- \ft12 R + \ft14 G_{ab} G^{ab} + \ft16 H_{abc} H^{abc} - L^{-2} N^2 - L^{-2} P_{a}P^{a} - V_{a}^{'kl} V^{'a}_{kl} \nn\\
&& \qquad  + L^{-1} \Box L - \ft12 L^{-2} \partial_a L \partial^a L \Big)  + 2L^{-1} P^a V'_{aij}  - \sqrt{2} L^{-1}\nabla^a(L V'_{a}{}^m{}_{(i} \d_{j)m}), \nn\\
\underline{\widehat{F}}_{ab}|_{\s=1} &=& 2\sqrt{2} \partial_{[a} \Big( V_{b]} + \sqrt2 L^{-1} P_{b]} \Big).
\eea
At this moment, we have also completed the vector multiplets coupled curvature squared terms. The general vector multiplets coupled $R+R^2$ theory is given by
\bea
 \cL_{LR}^D + \cL_{V}^{'D} + \cL_{\rm Riem^2}^{'D} + \cL_{C^2 + \ft16 R^2}^{'D} + \cL_{R^2}^{'D},
\eea
in which the vector multiplet action in $\s=1$ gauge is given as
\bea
e^{-1} \cL_{V}^{'D}|_{\s=1} &=& a_{IJ}\Big( \r Y_{ij}^I Y^{ij\,J} + 2\r^I Y_{ij}^J Y^{ij} + \r \r^I \nabla_\m \nabla^\m \r^J  + \ft12 \r \nabla_\m \r^I \nabla^\m \r^J \nn\\
&& \qquad -\ft14 \r ( F_{\m\n}^I - \r^I G_{\m\n}) ( F^{\m\n\,J} - 3 \r^J G^{\m\n}) -\ft12 (F_{\m\n}^I -  \r^I G_{\m\n} )\r^J F^{\m\n} \nn\\
&& \qquad + \ft1{12}  \r^I \r^J \e^{\m\n\r\s\l}(F_{\m\n} - 2\r G_{\m\n}) H_{\r\s\l} + \ft16 \r \r^I \e^{\m\n\r\s\l} F_{\m\n}^J H_{\r\s\l} \nn\\
&& \qquad -\ft18 \e^{\m\n\r\s\l} F_{\m\n}^I F_{\r\s}^J A_\l \Big) \,.
\label{YMS1}
\eea
The supersymmetric completion of vector multiplets coupled Gauss-Bonnet combination can be achieved by setting  $\g_I = 0$, and choosing the free parameters of $\cL_{\rm Riem^2}^{'D}$ and $\cL^{'D}_{C^2 +\ft16 R^2}$ to be related to each other according to $\b_I = 3 \a_I$.

\section{$D=5,\, \cN=2$ Off-Shell Supergravity in the Standard Weyl Multiplet}\label{section: Sugra}

In the previous section, we presented all off-shell curvature squared actions in five dimensional $\cN=2$ theory based on the dilaton Weyl multiplet. In this section, we focus on the construction of off-shell actions by using the standard Weyl multiplet.

In \cite{Bergshoeff:2004kh, Hanaki:2006pj}, a Poincar\'e supergravity was constructed by coupling the standard Weyl multiplet to a hypermultiplet and $n$ number of vector multiplets. However, we notice that to obtain the Ricci scalar squared action, it is more convenient to choose a linear multiplet as the compensator instead of a hypermultiplet. Therefore we devote this section to the construction of an off-shell Poincar\'e supergravity and its $R$-symmetry gauging using the linear and vector multiplets.

In subsection \ref{ss:GFPS1} we introduce the superconformal theory which give rise to an off-shell Poincar\'e supergravity upon fixing the redundant superconformal symmetries. In the next subsection, we discuss the $R$-symmetry gauging of this theory and show that the corresponding on-shell theory is the conventional gauged minimal $D=5,\, \cN=2$ supergravity.


\subsection{Poincar\'e Supergravity}\label{ss:GFPS1}

A consistent superconformal supergravity is given by combining the linear multiplet action and vector multiplet action
\bea
\cL_R^S &=& - \cL^S_L - 3 \cL_{V}^S \,,
\eea
where $\cL^S_L$ is given in (\ref{Lact}) and $\cL^S_V$ is given in (\ref{Vact}). This action has redundant superconformal symmetries needing be fixed in order to obtain an off-shell Poincar\'e supergravity. The gauge fixing conditions adopted in this section are
\bea
L_{ij} = \ft1{\sqrt2} \d_{ij},\qquad  b_\m = 0, \qquad \vf^i = 0,
\label{GF1}
\eea
where the first one breaks $SU(2)_R$ to $U(1)_R$ and fixes dilatation by setting $L=1$. The second one fixes the special conformal symmetry, and the last choice fixes the $S$-supersymmetry. To maintain the gauge (\ref{GF1}), the compensating transformations are required. Here we only present the compensating special supersymmetry and the compensating conformal boost with parameters
\bea
\eta_k &=& \ft13 \Big( \g \cdot T \e_k - \ft1{\sqrt2} N \d_{ik} \e^i + \ft1{\sqrt2} \rmi \slashed{E} \d_{ik} \e^i + \rmi \g^a V_a^{'(i}{}_l \d^{j)l} \d_{ik} \e_j \Big),  \label{ETA} \\
\L_{K\m} &=& - \ft14 \rmi \bar\e \phi_\m - \ft14 \rmi \bar\eta \p_\m + \bar\e \g_\m \chi. \label{LKM}
\eea
Using the gauge fixing conditions, the bosonic part of the corresponding off-shell Poincar\'e supergravity is given by
\bea
e^{-1} \cL_{R}^{S} &=& \ft18 (\cC+3)R + \ft13 (104 \cC - 8) T^2 + 4 (\cC-1)D - N^2- P_\m P^\m + V_\m^{'ij} V^{'\m}_{ij} - \sqrt2 V_\m P^\m \nn\\
&&+ \ft34 C_{IJK} \r^I F_{\m\n}^J F^{\m\n\, K}+ \ft32 C_{IJK} \r^I \partial_\m \r^J \partial^\m \r^K - 3 C_{IJK} \r^I Y_{ij}^J Y^{ij\,K}\nn\\
&& - 12 C_{IJK} \r^I \r^J F_{\m\n}^K T^{\m\n} + \ft18 \e^{\m\n\r\s\l} C_{IJK} A^I_\m F_{\n\r}^J F_{\s\l}^K,
\label{SWSUGRA}
\eea
where we have defined $\cC = C_{IJK} \r^I \r^J \r^K$. In the context of M-theory, the theory of five-dimensional ${\cal N} = 2$ supergravity coupled to Abelian vector supermultiplets arise by compactifying eleven-dimensional supergravity, the low-energy theory of M-theory, on a Calabi-Yau three-folds \cite{pap1,ant1}. $STU$ model corresponds to $\cC =STU$, where $S$, $T$ and $U$ are three vector moduli.


\subsection{Gauged Model}\label{ss:GM1}

As a result of our gauge choices (\ref{GF1}), the $\U(1)_R$ symmetry of the off-shell Poincare theory (\ref{SWSUGRA}) is gauged by the auxiliary vector $V_\m$, i.e. the full $\U(1)_R$ covariant derivative for gravitino is given by,
\bea
\nabla_\m \p_\n^i = \Big( \partial_\m + \ft14 \o_\m{}^{ab} \g_{ab} \Big) \p_\n^i - \ft12 V_\m \d^{ij} \p_{\n\,j} \,.
\label{auxcderiv}
\eea
where $V_{\m}^{'ij}$, the traceless part of $V_\m^{ij}$ does not appear in the $\U(1)_R$ covariant derivative for gravitino as a consequence of our gauge fixing choices (\ref{GF1}). In this section, we discuss the $\U(1)_R$ gauging of the Poincar\'e the theory by physical vectors $A_\m^I$. In the rest of the paper, we use the following notation
\bea
\cC = C_{IJK} \r^I \r^J \r^K, \quad \cC_I = 3 C_{IJK} \r^I \r^K,\quad \cC_{IJ} = 6 C_{IJK} \r^K\,.
\eea
The off-shell gauged model is given by
\bea
e^{-1}\cL_{g R}^{S} &=& e^{-1} ( \cL_{R}^S - 3 g_I \cL^I_{VL})|_{L=1}  \nn\\
&=& \ft18 (\cC + 3)R + \ft13 (104 \cC - 8) T^2 + 4 (\cC -1)D - N^2  - P_\m P^\m + V_\m^{'ij} V^{'\m}_{ij} \nn\\
&& - \sqrt2 P_\m V^\m + \ft18 \cC_{IJ} F_{\m\n}^I F^{\m\n\, J}  + \ft14 \cC_{IJ} \partial_\m \r^I \partial^\m \r^J - \ft12 \cC_{IJ} Y_{ij}^I Y^{ij\,J} - 4 \cC_{I}  F_{\m\n}^I T^{\m\n} \nn\\
&&  + \ft18 \e^{\m\n\r\s\l} C_{IJK} A^I_\m F_{\n\r}^J F_{\s\l}^K  - \ft3{\sqrt2} g_I Y^I_{ij} \d^{ij} - 3 g_I P^\m A_\m^I - 3 g_I \r^I N \,.
\eea
where $L=1$ indicates the gauge fixing condition (\ref{GF1}). As the field equation of $V_\m$ implies that $P_\m = 0$, we can immediately see that the $P_\m$ equation implies that
\bea
V_\m = - \ft{3}{\sqrt2} g_I A_\m^I  \,,
\label{Gauging}
\eea
hence, the auxiliary vector $V_\m$ is replaced by a linear combination of physical vectors $A_\m^I$
\bea
\nabla_\m \p_\n^i = \Big( \partial_\m + \ft14 \o_\m{}^{ab} \g_{ab} \Big) \p_\n^i + \ft3{2\sqrt2} g_I A^I_\m \d^{ij} \p_{\n\,j} \,.
\label{physgaug}
\eea
Therefore, the $\U(1)_R$ symmetry is gauged by the physical vectors. The equations of motion for $D, T_{ab}, N$ and $Y_{ij}^I$ lead to
\bea
0 &=& \cC - 1, \nn\\
0 &=& \ft23 (104 \cC - 8) T_{ab} - 4 \cC_I F_{ab}^I, \nn\\
0 &=& 2N + 3 g_I \r^I, \nn\\
0 &=& \cC_{IJ} Y_{ij}^J + \ft3{\sqrt2} g_I \d_{ij}.
\eea
The field equation for $D$ implies the constraint for very special geometry
\bea
C_{IJK} \r^I \r^J \r^K = 1.\,
\label{VSG}
\eea
Eliminating $T_{ab}$, $N$ and $Y_{ij}^I$ according to their field equations gives rise to
the following on-shell action
\bea
e^{-1}\cL_{g R}^{S}|_{\rm{on-shell}} &=& \ft12 R + \ft18 (\cC_{IJ} - \cC_I \cC_J) F_{\m\n}^I F^{\m\n\,J} + \ft14 \cC_{IJ} \partial^\m \r^I \partial^\m \r^J\nn\\
&&+ \ft18 \e^{\m\n\r\s\l} C_{IJK} A^I_\m F_{\n\r}^J F_{\s\l}^K + \Lambda(\rho),
\label{premin}
\eea
with
\bea
\L(\r) = \ft94 (g_I \r^I)^2 + \ft92 \cC^{IJ} g_I g_J,
\label{potential1}
\eea
where $\cC^{IJ}$ is the inverse of $\cC_{IJ}$ and due to the constraint (\ref{VSG}), $\r^I$ is not independent field now.
One can proceed further and truncate the on-shell vector multiplets to obtain the minimal gauged supergravity. In order to do so, we consider a single graviphoton via
\bea
\r^I=\bar{\r}^{I},\quad A_\m^I = \bar{\r}^I A_\m , \quad g_I = \bar{\r}_I g,
\label{truncation}
\eea
where $\bar{\rho}^I$ is VEV of the scalar at the critical value of the scalar potential (\ref{potential1}) and  $\bar{\r}_I$ satisfies $\bar{\rho}^I \bar{\rho}_I = 1$. The truncation conditions in (\ref{truncation}) are consistent with the supersymmetry transformation rules and lead to
\bea
e^{-1} \cL_{g R}^{{\rm min}} = \ft12 R - \ft38 F_{\m\n} F^{\m\n} + \ft18 \e^{\m\n\r\s\l} A_\m F_{\n\r} F_{\s\l} + 3 g^2,
\eea
which reproduces the conventional minimal on-shell supergravity in five dimensions.

\section{Supersymmetric Curvature Squared Actions in the Standard Weyl Multiplet}\label{section: Ricci2SW}

In this section, we quickly review the Weyl squared action derived in \cite{Hanaki:2006pj} and construct an off-shell Ricci scalar squared action based on the standard Weyl multiplet. The procedure for constructing the Ricci scalar squared action in the standard Weyl multiplet is similar to the one used in the dilaton Weyl multiplet. In the standard Weyl multiplet, the Ricci scalar squared term can be coupled to $n$ vector multiplets and alter the very special geometry.

\subsection{Supersymmetric Weyl Squared Action}
Using superconformal tensor calculus, an off-shell Weyl squared action in the standard Weyl multiplet was constructed in \cite{Hanaki:2006pj}, and its bosonic part reads
\bea
e^{-1} \cL_{C^2}^S &=&  c_{I} \Big[ \ft18\r^I {C}^{\m\n\r\s} {C}_{\m\n\r\s}+ \ft{64}3 \r^I {D}^2 + \ft{1024}9 \r^I {T}^2 {D} - \ft{32}3 {D} \, {T_{\m\n}} F^{\m\n\,I}   \nn\\
&&  - \ft{16}3 \r^I {C}_{\m\n\r\s} {T}^{\m\n} \, {T}^{\r\s} + 2{C}_{\m\n\r\s} {T}^{\m\n} F^{\r\s\,I} + \ft1{16} \e^{\m\n\r\s\l}A_\m^I {C}_{\n\r\t\d} {C}_{\s\l}{}^{\t\d}    \nn\\
&& -\ft1{12} \e^{\m\n\r\s\l} A_\m^I {V}_{\n\r}{}^{ij} {V}_{\s\l\, ij} +  \ft{16}3 Y^I_{ij} {V}_{\m\n}{}^ {ij} {T}^{\m\n} - \ft{1}3 \r^I {V}_{\m\n}{}^{ij}{V}^{\m\n}{}_{ij} \nn\\
&& +\ft{64}3 \r^I  \nabla_\n T_{\m\r} \nabla^\m T^{\n\r} - \ft{128}3 \r^I {T_{\m\n}} \nabla^\n \nabla_\r {T}^{\m\r} - \ft{256}9 \r^I R^{\n\r} T_{\m\n} T^\m{}_\r \nn\\
&& + \ft{32}9 \r^I R T^2 - \ft{64}3  \r^I \nabla_\m T_{\n\r}\nabla^\m T^{\n\r} + 1024 \r^I \, {T}^4- \ft{2816}{27} \r^I  ({T}^2)^2   \nn\\
&&- \ft{64}9 {T_{\m\n}} F^{\m\n\,I} {T}^2 - \ft{256}3 {T_{\m\r}} {T}^{\r\l} {T_{\n\l}} F^{\m\n\,I}  - \ft{32}3   \e_{\m\n\r\s\l}  {T}^{\r\t} \nabla_\t {T}^{\s\l} F^{\m\n\,I}  \nn\\
&& - 16   \e_{\m\n\r\s\l} {T}^\rho{}_\t \nabla^\s {T}^{\l\t} F^{\m\n\,I}  - \ft{128}3 \r^I \e_{\m\n\r\s\l} {T}^{\m\n} {T}^{\r\s} \nabla_\t {T}^{\l\t}\Big] \,,
\label{pregb}
\eea
where the five dimensional Weyl tensor reads
\bea
C_{\m\n\r\s} &=& R_{\m\n\r\s} - \ft13 (g_{\m\r} R_{\n\s} - g_{\n\r} R_{\m\s} - g_{\m\s} R_{\n\r} + g_{\n\s} R_{\m\r}) \nn\\
&& + \ft{1}{12} (g_{\m\r} g_{\n\s} - g_{\m\s} g_{\n\r}) R.
\eea
Now, we would like to comment more rigorously on the difference between the Weyl squared invariant in the standard Weyl multiplet (\ref{pregb}) and its counterpart in the dilaton Weyl multiplet (\ref{pregb2}). As mentioned before, one of the main differences between these actions relies on the definition of $D$ which is an independent field in the standard multiplet but a composite field in the dilaton Weyl multiplet. As a composite field in the dilaton Weyl multiplet, $D$ contains a curvature term (\ref{UMap}). However, simply replacing $D, T_{ab}$ and $\chi^i$ by their composite expressions does not produce an action solely based on the dilaton Weyl multiplet. The resulting action also depends on the fields in the vector multiplet.  We recall that neither two-derivative Poincar\'e supergravity (\ref{RLag}) nor the Riemann squared action (\ref{bosonicR2}) has any dependence on the vector multiplet in the minimal off-shell supersymmetric model. To obtain the Weyl squared actions solely constructed in terms of the dilaton Weyl multiplet, the map (\ref{DVMap}) from the dilaton Weyl multiplet to the vector multiplet is indispensable.

The Weyl squared action in (\ref{pregb}) is invariant under the transformation rules given in (\ref{SWMTR}) and (\ref{VMTR}) with $\eta^i$ and $\Lambda_{K\mu}$ being replaced according to (\ref{ETA}) and (\ref{LKM}).
\subsection{Supersymmetric Ricci Scalar Squared Action}
To obtain the Ricci scalar squared invariant in the standard Weyl multiplet, we begin with the composite expressions given in (\ref{LEmbde}) after fixing the redundant symmetries,
\bea
\underline{\r}|_{L=1} &=& 2N,  \nn\\
\underline{Y}^{ij}|_{L=1} &=&  \ft1{\sqrt2} \d^{ij}\Big(- \ft3{8} R -  N^2 - P^2 + \ft83 T^2 + 4D - V_{a}^{'kl} V_{kl}^{'a} \Big)   \nn\\
&& + 2 P^a V'_{a}{}^{ij}  - \sqrt2 \nabla^a V'_{a}{}^{m(i} \d^{j)}{}_{m} , \nn\\
\underline{F}^{ab}|_{L=1} &=& 2\sqrt{2} \partial^{[a} \Big( V^{b]} + \sqrt2 P^{b]} \Big).
\label{fixedmap}
\eea
From the above expressions, one sees that the Ricci scalar squared can come from $Y_{ij} Y^{ij}$ term in the vector action. By choosing $C_{I11} = a_I$ and all other possibilities to zero in (\ref{Vact}), we obtain the following Ricci scalar squared action
\bea
e^{-1} \cL_{R^2}^{ S} &=& a_I \Big( \r^I \underline{Y}_{ij} \underline{Y}^{ij} + 2 \underline{\r} \underline{Y}^{ij} Y_{ij}^I-\ft18\r^I \underline{\r}^2 R - \ft14 \r^I  \underline{F}_{\m\n}  \underline{F}^{\m\n} - \ft12 \underline{\r} \, \underline{F}^{\m\n} F_{\m\n}^I  \nn\\
&& \quad+ \ft12 \r^I \partial_\m \underline{\r} \partial^\m \underline{\r} + \r^I \underline{\r} \Box \underline{\r} - 4 \r^I \underline{\r}^2 (D + \ft{26}3 T^2) + 4 \underline{\r}^2 F_{\m\n}^I T^{\m\n} \nn\\
&& \quad + 8 \r^I \underline{\r}\, \underline{F}_{\m\n} T^{\m\n} - \ft18 \e_{\m\n\r\s\l} A^{\m I} \underline{F}^{\n\r} \underline{F}^{\s\l} \Big).
\label{R2SW}
\eea
As we will extensively work on this action in the following section, here we postpone to plug in the composite expressions given in (\ref{fixedmap}). To study the solutions with vanishing auxiliary fields, we can use a truncated version of above action as we shall see in the next section. Off-shell supersymmetry allows us to combine the Ricci scalar squared action with the two-derivative Poincar\'e supergravity (\ref{SWSUGRA}) and the Weyl squared action (\ref{pregb}) to form a more general supergravity theory
\bea
\cL_R^{S} + \cL_{C^2}^{S} + \cL_{R^2}^{S},
\eea
where $\cL_R^S$ is given in (\ref{SWSUGRA}), $\cL_{C^2}^S$ is given in (\ref{pregb}) and $\cL_{R^2}^S$ is given in (\ref{R2SW}).

\subsection{Ricci Scalar Squared Extended Gauged Model and Corrected Very Special Geometry}
In this section, we consider the off-shell Ricci scalar squared extended gauged model
\bea
\cL_R^{S} + \cL_{R^2}^{S} + g_I \cL_{VL}^I.
\label{gextended}
\eea
We consider the maximal supersymmetric $AdS_5$ solutions. The ansatz preserving $SO(4,2)$ symmetry takes the form
\be
R_{\m\n\r\l}=\ft{R}{20}(g_{\m\r}g_{\n\l}-g_{\m\l}g_{\n\r}),\quad A^{I}_{\m}=0,\quad T_{\m\n}=0,\quad \r=\bar{\r},\quad N=\bar{N},\quad D=\bar{D},
\ee
where $\bar{\r}$, $\bar{N}$ and $\bar{D}$ are some constants.
The maximal supersymmetry requires that
\be
R=-\ft{40}9\bar{N}^2,\quad Y_{ij}^I = \ft1{3\sqrt2} \bar{ \r}^I \bar{N} \d_{ij}, \quad \bar{D}=0.
\label{Maximal}
\ee
Employing $\r^{I}$ equation and $N$ equation for the Lagrangian (\ref{gextended}), we obtain
\be
2 \bar{N} C_{IJK} \bar{\r}^J \bar{\r}^K +3 g_I - \ft83 a_I \bar{N}^3=0 ,\quad 2 \bar{N} + 3 g_I \bar{\r}^I = 0.
\ee
These two equations imply
\be
\bar{C} = 1 + \ft43 a_I \bar{\r}^I \bar{N}^2,
\label{qcvsg}
\ee
which is consistent with $D$ field equation, $Y^{Iij}$ equation and Einstein equation. Therefore, in the presence of Ricci scalar squared term, $AdS_5$ maintains to be the maximally supersymmetric solution. However, in this case, the very special geometry is modified according to (\ref{qcvsg}). Inserting $N=-\ft32g_I \bar{\r}^I$ into (\ref{qcvsg}), the quantum corrected very special geometry on the moduli space of $AdS_5$ vacuum can be written as
\be
\widetilde{C}_{IJK}\bar{\r}^I\bar{\r}^J\bar{\r}^k=1,\qquad \widetilde{C}_{IJK}=C_{IJK}+ 3 a_{(I}g_{J}g_{K)}.
\ee
We emphasize that the inclusion of the Weyl squared action ({\ref{pregb}) also modifies the definition of very special geometry, however the modification vanishes on the maximally supersymmetric $AdS_5$ background (\ref{Maximal}).

\section{Supersymmeric Solutions with $AdS_3 \times S^2$ and $AdS_2 \times S^3$ Near Horizon Geometries} \label{section: Solution}
The strategy for finding regular solutions in higher derivative theory is to first write an ansatz consistent
with the assumed symmetries, and then demand unbroken supersymmetry. The supersymmetric magnetic strings and electric black holes preserving one half of the supersymmetries have been studied in \cite{sab1,sab2} for the case
of $n$ vector multiplets coupled to two-derivative Poincar\'e supergravity and in \cite{cas1} for the higher derivative case where only the off-shell Weyl squared invariant is taken into account. In the presence of Weyl squared, the magnetic strings and electric black holes receive corrections. In the following, we consider the Ricci scalar squared extended two-derivative theory (\ref{SWSUGRA}, which is the simplest curvature squared extended model. Explicitly, in this section we study the theory
\bea
\cL = \cL_{R}^{S} + \cL_{R^2}^{S},
\label{R2Ext}
\eea
where $\cL_{R}^{S}$ and $\cL_{R^2}^{S}$ are given by (\ref{SWSUGRA}) and (\ref{R2SW}) respectively. We are interested in solutions with vanishing auxiliary fields. It can be checked that
\bea
Y_{ij}^I = N = E_a = V_a = V_{a}^{'ij} = 0,
\eea
is a consistent truncation of (\ref{R2Ext}) leading to a simpler effective action describing the
very special geometry extended by Ricci scalar squared invariant
\bea
e^{-1} \cL &=& \ft18 ({\cal C}+3)R + \ft13 (104 {\cal C} - 8) T^2 + 4 ({\cal C}-1)D + \ft34 C_{IJK} \r^I F_{ab}^J F^{ab\, K}    \nn\\
&& + \ft32 C_{IJK} \r^I \partial_\m \r^J \partial^\m \r^K - 12 C_{IJK} \r^I \r^J F_{ab}^K T^{ab} + \ft18 \e^{abcde} C_{IJK} A^I_a F_{bc}^J F_{de}^K \nn\\
&& + a_I \r^I \Big( \ft{9}{64} R^2 - 3 D R - 2 R T^2 + 16 D^2 + \ft{64}3 D T^2 + \ft{64}9 (T^2)^2 \Big).
\label{R+nV+R2}
\eea
The supersymmetry transformations for the fermionic fields take the following forms when the auxiliary fields vanish
\bea
\d \p_\m^i &=& \Big( \nabla_\m - 4 \rmi \g^a T_{\m a} + \ft23 \rmi \g_\m \g \cdot T \Big) \e^i, \nn\\
\d \chi^i &=& \Big( \ft14 D + \ft18 \rmi \g^{ab} \slashed{\nabla} T_{ab} - \ft18 \rmi \g^a \nabla^b T_{ab} - \ft16 \g^{abcd} T_{ab} T_{cd} \Big) \e^i, \nn\\
\d \l_i^I &=& \Big( -\ft14 \g \cdot \widehat{F}^I - \ft12 \rmi \slashed{\nabla} \r^I + \ft43 \r^I \g \cdot T \Big) \e_i.
\eea

\subsection{Magnetic string solutions}

The metric preserving the symmetry of a static string takes the form
\be
ds^2 = e^{2U_1(r)} (- dt^2+  dx_1^2) + e^{-4U_2(r)} dx^i dx^i,\quad dx^i dx^i = dr^2 + r^2 d\O_2^2,
\label{ms}
\ee
where $i=2,3,4$. $F^I_{ab}$ and $T_{ab}$ are chosen to be proportional to the volume form of $S^2$. A natural choice for the veilbein is given by
\be
e^{\hat{a}} = e^{U_1} dx^a , \quad a= 0,1, \qquad e^{\hat{i}}=e^{-2 U_2} dx^i , \quad i = 2,3,4.
\ee
Accordingly, the non-vanishing components of the spin connections are
\be
\o_a{}^{\hat{a}\hat{i}} =e^{U_1 + 2 U_2} \partial_i U_1, \quad
\o_k{}^{\hat{i}\hat{j}} = - 2 \d_k^i \partial_j U_2 + 2 \d_k^j \partial_j U_2.
\ee
Similar to \cite{cas1}, the supersymmmetry parameter $\epsilon^i$ is constant along the string and obeys
the projection condition which breaks half of the supersymmetries
\be
  \g_{\hat{t}\hat{1}}\e = -\e.
\label{pro}
\ee
We first study the gravitino variation which fixes $U_1=U_2$.
\be
\d \p_\m  = \Big( \nabla_\m - 4 \rmi \g^a T_{\m a} + \ft23 \rmi \g_\m \g \cdot T \Big) \e.
\ee
The covariant derivative is
\bea
\nabla_a &=& \partial_a + \ft12 e^{U_1 + 2 U_2} \partial_i U_1 \g_{\hat{a}\hat{i}}, \nn\\
\nabla_i &=& \partial_i + \partial_j U_2 \g_{\hat{j}\hat{i}}.
\eea
Along the string direction, we have
\bea
\Big[ \ft12 e^{U_1 + 2 U_2} \partial_i U_1 \g_{\hat{a}\hat{i}} + \ft23 \rmi e^{U_1} \g_{\hat{a}\hat{i}\hat{j}}, T^{\hat{i}\hat{j}} \Big] \e = 0.
\eea
We use the convention that $\g^0\g^1\g^2\g^3\g^4=i\epsilon^{01234}$ with $\epsilon^{01234}=1$. Therefore (\ref{pro}) implies
\be
\g_{\hat{i}\hat{j}\hat{k}} \e = \rmi \, \e_{\hat{i}\hat{j}\hat{k}} \e,\quad \g_{\hat{i}\hat{j}} \e = \e_{\hat{i}\hat{j}\hat{k}} \g_{\hat{k}} \e.
\ee
where $\e_{234} = 1$.
Using the above conditions, it can be obtained that
\bea
\Big[ \ft12 e^{U_1 + 2 U_2} \partial_k U_1 - \ft23 e^{U_1} T^{\hat{i}\hat{j}} \e_{\hat{i}\hat{j}\hat{k}} \Big] \g_{\hat{a}\hat{k}} \e = 0.
\eea
The auxiliary field $T_{ab}$ can be solved as
\bea
T_{\hat{i}\hat{j}} &=& \ft38 e^{2U_2} \e_{\hat{i}\hat{j}\hat{k}} \partial_k U_1.
\eea
The gravitino variatoin along $x^i$ direction leads to
\bea
\Big[ \partial_k - \rmi \, \e_{\hat{i}\hat{j}\hat{k}} \partial_{i} U_2 \g_{\hat{j}} -\ft83 \rmi \, \g^{\hat{i}} T_{k\hat{i}} - \ft23 \e_{\hat{i}\hat{j}\hat{l}} \, e^{\hat{l}}{}_{k} T_{\hat{i}\hat{j}} \Big] \e &=& 0.
\eea
The ``radial" part and ``angular" part result in two conditions
\bea
0 &=& \Big[ \partial_k - \ft23 \e_{\hat{i}\hat{j}\hat{l}} e{}^{\hat{l}}{}_k T_{\hat{i}\hat{j}} \Big] \e \,,\nn\\
0 &=& \Big[ -\e_{\hat{i}\hat{k}\hat{j}} \partial_i U_2 + \ft83 T_{k\hat{j}} \Big] \g_{\hat{j}} \e.
\eea
The second equation restricts
\be
U_1 = U_2,
\ee
then the first equation implies Killing spinor is
\be
\e = e^{U/2} \e_0,
\ee
where $\e_0$ is some constant spinor.
In cylindrical coordinates, $T_{ab}$ can be expressed as
\be
T_{\theta\phi} = \ft38 e^{-2U} r^2 \sin\theta \partial_r U, \quad T_{\hat{\theta} \hat{\phi}} = \ft38 e^{2U} \partial_r U.
\ee
The projection in cylindrical coordinates can be written as
\be
\g_{\hat{r}\hat{\theta}\hat{\phi}} \, \e = \rmi \, \e.
\ee

The gaugino variation $\d \l_i^I$ on the magnetic background gives
\be
\Big( -\ft12 \gamma_{\hat{\theta}\hat{\phi}} F^{I\, \hat{\theta}\hat{\phi}} - \ft12 \rmi \g^{\hat{r}} e_{\hat{r}}{}^r \partial_{r} \r^I + \ft83 \r^I \gamma_{\hat{\theta}\hat{\phi}} T^{\hat{\theta}\hat{\phi}} \Big) \e = 0.
\ee
Then
\be
F_{\hat{\theta}\hat{\phi}}^I =  -e^{2U_1} \partial_r \rho^I + \ft{16}3 \r^I T_{\hat{\theta}\hat{\phi}}=- \partial_r (\r^I e^{-2U}) e^{4U}.
\ee

The supersymmetry variation of $\chi^i$ leads to
\be
\Big( \ft14D + \ft18 \rmi \g^{ab} \slashed{\nabla} T_{ab} - \ft18 \rmi \g^a \nabla^b T_{ab} - \ft16 \g^{abcd} T_{ab} T_{cd} \Big) \e = 0.
\ee
Explicit computation shows that auxiliary field $D$ can be solved from the above equation as
\be
D = \ft38 e^{6U} r^{-2} \partial_r (e^{-3U} r^2 \partial_r U )= -\ft3{16} e^{6U} \nabla^2 e^{-2U}.
\ee

So far we have exhausted the constraints which can be derived from the variations of fermions.
In the following, we have to use the equations of motion. For the magnetic string configuration,
the equations of motion of gauge potential are automatically satisfied, however, the Bianchi identity
gives rise to
\be
\partial_r F_{\theta\phi}^I = - \partial_r \Big( r^2 \partial_r (\r^I e^{-2U}) \Big) \sin\theta = 0.
\ee
The solution to the above equation is given by \cite{sab1}
\be
\r^I e^{-2U} = H^I = \r^I_{\infty} + \frac{p^I}{2r},\quad F^I =  \frac{p^I}2 \e_2,
\ee
where $\r^I_\infty$ is the value of $\r^I$ in the asymptotically flat region where $U=0$.

The equation of $D$ derived from action (\ref{R+nV+R2}) is
\be
{\cal C} = 1 +a_I \r^I \Big( \ft34 R - 8 D - \ft{16}3 T^2 \Big).
\ee
After substituting $T_{ab}$, $D$ and $R$ according to
\be
T_{\hat{\theta} \hat{\phi}} = \ft38 e^{2U} \partial_r U,\quad D=-\ft3{16} e^{6U} \nabla^2 e^{-2U},\quad R = \frac{2 e^{4U}}r (4 U' - 3 r U^{'2} + 2 r U''),
\ee
where ``prime'' means derivative with respect to $r$. We find that the higher derivative corrections to the $D$ equation of motion vanishes. Similarly, there are no
higher derivative corrections to the equations of motion of $T_{ab}$, $g_{\m\n}$ and $\rho^I$. Therefore, the magnetic strings do not receive corrections from the Ricci scalar squared invariant. This result seems to be compatible with the expectation from string theory. From string theory point of view, the Ricci scalar squared invariant has no effects on the on-shell quantities since it can just come from a field redefinition of the two
derivative action. This result also suggests that it is the supersymmetrization of curvature squared terms
which captures the correct feature of quantum corrections of ${\cal N}=2$ string vacua, because an arbitrary combination of $R^2$, $D$ and $T_{ab}$ will modify the equations of motion in general.

\subsection{Electric black holes}
Finding electric black holes follow the procedure as \cite{cas1}.
Given the ansatz
\be
ds^2 = -e^{4U_1(r)} dt^2+   e^{-2U_2(r)} dx^i dx^i,\quad dx^i dx^i = dr^2 + r^2 d\O_3^2.
\label{eb}
\ee
Supersymmetry requires that
\be
U_1=U_2, \quad T_{ti} = \ft38 e^{2U} \partial_i  U , \quad A_{t}^I = -e^{2U} \rho^I\,,\quad D = \ft3{16} e^{2U} ( 3 r^{-1} U' + U'' - 2 r^{-2} U^{'2}).
\ee
In this case,
\be
R = \frac{2 e^{2U}}{r} (3 U' - 3 r U^{'2} +r U'').
\ee
Again, it can be checked that the higher derivative corrections to the equations of motion vanish. Therefore,
the electric black holes are not modified by the inclusion of Ricci scalar invariant.


\section{Conclusions}\label{section: conc}

In this paper, using superconformal tensor calculus, we have completed all off-shell curvature squared invariants in $D=5,\, \cN=2$ supergravity based on the dilaton Weyl multiplet for both the minimal and the vector multiplets coupled curvature squared invariants, namely the complete minimal curvature squared invariants consist of
\bea
\alpha \cL_{{\rm Riem}^2}^D + \beta \cL^D_{C^2 + \ft16 R^2} + \gamma \cL_{R^2}^D \,,
\eea
and the complete vector multiplets coupled curvature squared invariants take the form
\bea
\cL_{{\rm Riem}^2}^{'D} + \cL^{'D}_{C^2 + \ft16 R^2} + \cL_{R^2}^{'D} \,.
\eea
 Adopting the standard Weyl multiplet, we also constructed an off-shell Poincar\'e supergravity by using the linear and vector multiplets as compensators and a supersymmetric Ricci scalar squared coupled to $n$ vector multiplets. In the standard Weyl multiplet, the curvature squared extended model is generalized to take the form
\bea
\cL_{R}^S + \cL^S_{C^2} + \cL_{R^2}^S \,.
\eea
It is known that the gauged two-derivative Poincar\'e supergravity possesses an maximally supersymmetric $AdS_5$ vacuum solution. When the Ricci scalar squared is included, the very special geometry defined on the moduli space gets modified in a very simple way. Finally, we study the effects of Ricci scalar squared to the supersymmetric magnetic string and electric black hole solutions which are the 1/2 BPS solutions of the ungauged two-derivative theory. It is found that neither the magnetic string nor the electric black hole solutions gets modified by the supersymmetric completion of the Ricci scalar squared.

A comparison between the results in the dilaton Weyl multiplet and the standard Weyl multiplet leads to a natural question that what is the analogue of supersymmetric Riemann squared in the standard Weyl multiplet. At this moment, we do not know the exact answer. However, if such an invariant exists, one should be able to recover the Riemann squared invariant based on the dilaton Weyl multiplet from the Riemann squared invariant based on the standard Weyl multiplet by using the map (\ref{DVMap}). This argument constrains the form of the supersymmetric Riemann squared in the standard Weyl multiplet.

The modified very special geometry around the $AdS_5$ vacuum by the Ricci scalar squared invariant is very intriguing unlike the supersymmetric completion of the Weyl tensor squared which does not affect the definition of very special geometry in the $AdS_5$ vacuum. Interpretation of the modified very special geometry from string/M theory and its application in the context of AdS/CFT correspondence deserve future investigation. Finally, our procedure for the construction of Ricci scalar squared invariant can be straightforwardly generalized to $D=6, \, \cN =(1,0)$ supergravity \cite{op2}.

\section*{Acknowledgements}
We thank Ergin Sezgin for discussions and Bernard de Wit for bringing the interesting paper \cite{deWit:2006gn} to us. Y.P. is supported in part by DOE grant DE-FG03-95ER40917. Y.P. would also like to thank the hospitality of CHEP at Peking University and Department of Physics and Astronomy at Shanghai Jiao Tong University where this work is completed.

\appendix
\section{Composite Expressions for Linear and Vector Multiplets}

The elements of linear multiplet can be written in terms of the elements of a vector multiplet and a Weyl multiplet \cite{Fujita:2001kv}
\bea
L_{ij} &=& 2 \rho Y_{ij} - \ft12 \rmi \bar\l_{i} \l_{j}, \nn\\
\vf_i  &=& \rmi \r \slashed{\cD}\l_i + 2 \r \g \cdot T \l_i - 8 \r^2 \chi_i - \ft14 \g \cdot \widehat{F} \l_i + \ft12 \slashed{\cD}\r \l_i - Y_{ij} \l^j,\nn\\
E^a &=& \cD_b \Big( -\r F^{ab} + 8 \r^2 T^{ab} - \ft14 \rmi \bar\l \g_{ab} \l \Big) - \ft18 \e^{abcde} F_{bc} F_{de}, \nn\\
N &=& \r \Box^C \r + \ft12 \cD_a \r \cD^a \r - \ft14 \widehat{F}_{ab} \widehat{F}^{ab}+ Y^{ij} Y_{ij}+ 8 \widehat{F}_{ab} T^{ab}  \nn\\
&& - 4 \r^2 \Big(D + \ft{26}3 T^2\Big)  - \ft12 \bar\l \slashed{\cD} \l + \rmi \bar\l \g\cdot T \l + 16 \rmi \r \bar\chi \l.
\label{VEmbed}
\eea
Using these composite expressions in the vector-linear action (\ref{densityformula}), one obtains the vector multiplet action given in (\ref{Vact}).

One can also construct the elements of vector multiplet in terms of the elements of a linear multiplet and a Weyl multiplet \cite{Coomans:2012cf, Ozkan:2013uk}
\bea
\r &=& 2 L^{-1} N + \rmi L^{-3} \bar\vf^i \vf^j L_{ij} \,, \nn\\
\l_i &=& -2 \rmi \slashed{\cal{D}}\vf_i L^{-1} + ( 16 L_{ij}\chi^j + 4 \g \cdot T \vf_i) L^{-1}  - 2 N L_{ij} \vf^j L^{-3}  \nn \\
&& + 2\rmi (\slashed{\cD} L_{ij} L^{jk} \vf_k - \slashed{E} L_{ij} \vf^j ) L^{-3}  + 2 \rmi \vf^j \bar\vf_i \vf_j L^{-3} - 6 \rmi \vf^j \bar\vf_k \vf_l L^{kl} L_{ij} L^{-5} ,\nn\\
Y_{ij} &=& L^{-1} \Box^C L_{ij} - {\cal{D}}_{a}L_{k(i} {\cal{D}}^{a}L_{j)m} L^{km} L^{-3} - N^2 L_{ij} L^{-3} - E_{\m}E^{\m}L_{ij} L^{-3}\nn\\
&& + \tfrac{8}{3} L^{-1} T^2 L_{ij} + 4 L^{-1} D L_{ij} + 2 E_{\m} L_{k(i} {\cal{D}}^{\m} L_{j)}{}^{k} L^{-3}  -\ft12 \rmi N L^{-3} \bar\vf_{(i} \vf_{j)} \nn\\
&& - \ft43 L^{-5}N  L_{k(i} L_{j)m} \bar\vf^k \vf^m - \ft23 L^{-3} \bar\vf_{(i} \slashed{E} \vf_{j)} - \ft13 L^{-5} L_{k(i} L_{j)m} \bar\vf^k \slashed{E} \vf^m \nn\\
&& - 8 \rmi   L^{-1} \bar\chi _{(i} \vf_{j)} + 16 \rmi  L^{-3} L_{k(i} L_{j)m} \bar\chi^k \vf^m +2 L^{-3} L_{k(i} \bar\vf^k \slashed{\cD}\vf_{j)} \nn\\
&& + 2 \rmi L^{-3} L_{ij} \bar\vf \g \cdot T \vf - \ft23 L^{-3} \bar\vf_{(i} \slashed{\cD} L_{j)k} \vf^k - L^{-5} L_{mn} L^k{}_{(i} \bar\vf^m \slashed{\cD}L_{j)k}\vf^n \nn\\
&& -\ft16  L^{-5} L_{km} \bar\vf_i \g^a \vf_j \bar\vf^k \g_a \vf^m +\ft1{12} L^{-7} L_{ij} L_{km} L^{pq} \bar\vf^k \g_a \vf^m \bar\vf_p \g^a \vf_q ,\nn\\
\widehat{F}_{\m\n} &=& 4 {\cal{D}}_{[\m}(L^{-1} E_{\n ]}) + 2L^{-1} \widehat{R}_{\m\n}{}^{ij}(V) L_{ij} - 2 L^{-3} L_{k}^{l} {\cal{D}}_{[\m}L^{kp} {\cal{D}}_{\n ]} L_{lp}  \nn\\
&& - 2 \cD_{[\m} (L^{-3} \bar\vf^i \g_{\n]} \vf^j L_{ij} ) - \rmi L^{-1} \bar\vf \widehat{R}_{\m\n} (Q) \,.
\label{LEmbde}
\eea
The composite expressions for the elements of vector multiplet can be used in the density formula (\ref{densityformula}) to give rise to a linear multiplet action (\ref{Lact}), or in vector multiplet actions (\ref{Vact}), (\ref{VDW}) to obtain Ricci scalar squared invariants.

\section{Comparison Between Different Notations}\label{section: notation}

\begin{center}
\begin{tabular}{ c | c }
Hanaki et. al. & Bergshoeff et. al  \\
\hline \hline
$\eta_{\m\n}$ & $- \eta_{\m\n}$ \\
$x_\m$ & $x^\m$ or $-x_\m$ \\
$\partial_\m$ & $\partial_\m$ or $-\partial^\m$ \\
$\g^\m$ & $\rmi \g^\m$ or $-\rmi \g_\m$ \\
$\g^\m \partial_\m = \slashed{\partial}$ & $\rmi \g^\m \partial_\m = \rmi \slashed{\partial}$ \\
\hline \hline
\multicolumn{2}{c}{Transformation Parameters} \\
\hline \hline
$\e^i$ & $\ft12 \e^i$ \\
$\eta^i$ & $-\ft12 \eta^i - \ft12 \g \cdot T \e^i$ \\
$\xi_{K\m}$&$-\L_{K\m} + \bar\epsilon \g_\m \chi$ \\
\hline \hline
\multicolumn{2}{c}{Standard Weyl Multiplet} \\
\hline\hline
$e_\m{}^a$ & $e_\m{}^a$  \\
$\p_\m^i$ & $\ft12\p_\m^i$ \\
$\phi_\m^i$ & $-\ft12 \phi_\m^i - \ft12 \g \cdot T \p_\m^i$\\
$\o_\m{}^{ab}$ & $\o_\m{}^{ab}$\\
$b_\m$ & $b_\m$\\
$V_\m^{ij}$ & $-V_\m^{ij}$\\
$v_{ab}$ & $4T_{ab}$ \\
$\chi^i$ & $32 \chi^i$ \\
$D$ & $16D + \ft{128}3 T^2$\\
\hline \hline
\multicolumn{2}{c}{Vector Multiplet} \\
\hline\hline
$W_\m$ & $A_\m$ \\
$\O^i$ & $-\ft12 \l^i$ \\
$M$ & $ - \r$ \\
$Y^{ij}$ & $- Y^{ij}$ \\
\hline \hline
\multicolumn{2}{c}{Linear Multiplet} \\
\hline\hline
$L_{ij}$ & $L_{ij}$ \\
$\vf^i$ & $\vf^i$ \\
$E_a$ & $-2E_a$ \\
$N$ & $2N$\\
\hline \hline
\end{tabular}
\end{center}

\end{document}